\documentclass[twocolumn,showpacs,preprintnumbers,amsmath,amssymb]{revtex4-1}
\usepackage{graphicx}
\usepackage{dcolumn}
\usepackage{bm}
\usepackage{color}
\usepackage{braket} 

\newcommand{\be}{\begin{eqnarray}}
\newcommand{\ee}{\end{eqnarray}}

\begin{document}


\title{Application of the inhomogeneous Kibble-Zurek mechanism to quench dynamics in the transition from a Mott-insulator to a superfluid in a finite system}

\author{Yoshihiro Machida}
\author{Kenichi Kasamatsu}
\affiliation{
Department of Physics, Kindai University, Higashi-Osaka, Osaka 577-8502, Japan}

\date{\today}

\begin{abstract}
We apply the theory of inhomogeneous Kibble-Zurek mechanism to understand quench dynamics from the Mott insulator to the 
superfluid in a cold Bose gases confined in both a two-dimensional optical lattice and a harmonic trap. 
The local quench time and the resulting Kibble-Zurek diabatic-adiabatic boundary take a nontrivial positional dependence due to the Mott-lobe 
structure of the ground state phase diagram of the Bose-Hubbard model. We demonstrate the quench dynamics through the time-dependent Gutzwiller simulations, revealing inhomogeneous properties of the growth of the superfluid order parameter. 
The inhomogeneous Kibble-Zurek theory is applicable for the shallow harmonic trap.  
\end{abstract}

\pacs{
03.75.Kk, 
47.20.Ft, 
67.85.Fg 
} 
\maketitle

\section{Introduction} \label{intro}
Ultracold atomic gases are versatile testing beds for studying many-body quantum phenomena under isolated, clean, and highly controllable environments \cite{lewenstein2012ultracold}. 
In this system, nonequilibrium quantum dynamics under a sudden quench 
of the system parameters, called quantum quench, is one of the major topics, providing many challenging problems of many-body quantum dynamics that 
are difficult to be solved by existing theoretical treatments \cite{polkovnikov2011colloquium,kennett2013out}. 

The Kibble-Zurek mechanism (KZM) is a well-known theory that describes the nonequilibrium process of the 
phase transition from a symmetric phase to a symmetry-breaking one, predicting a density of topological defects generated by a
rapid quench of the parameters which induce the phase transition \cite{kibble1976topology,zurek1985cosmological}. 
The KZM has been studied for decades in various condensed matter systems \cite{zurek1996cosmological,campo2014universality}. 
Later, the theory has been extended to quantum phase transitions \cite{dziarmaga2010dynamics} and investigated in cold Bose gases \cite{sadler2006spontaneous,chen2011quantum,lamporesi2013spontaneous,braun2015emergence,navon2015critical,anquez2016quantum,chen2019dynamical}. 
In the case of a Bose gas in an optical lattice, the Mott insulator (MI) and the superfluid (SF) can exist as the ground state, which depends on a depth of an optical lattice, strength of interatomic interactions, and particle fillings \cite{jaksch1998cold,greiner2002quantum}. 
The experiments of Refs.~\cite{chen2011quantum,braun2015emergence} reported the quench dynamics from the MI to the SF phase, 
where the applicability of the KZM has been also discussed. 
The theoretical works of the KZM in the MI-SF transition has been studied by several 
authors \cite{cucchietti2007dynamics,horiguchi2009non,dziarmaga2012quench,shimizu2018dynamics,weiss2018kibble,zhou2020quench}. 

In this work, we consider the quench dynamics of the Bose-Hubbard model (BHM) in the presence of a harmonic confinement. 
The impact of the harmonic potential, which makes the system inhomogeneous, to the KZM in the MI-SF transition has not been considered seriously so far. 
The harmonic potential gives rise to the spatially dependent chemical potential, resulting in the core-shell structure of the SF and MI domains 
\cite{jaksch1998cold,batrouni2002mott}. 
The inhomogeneous nature of the equilibration dynamics from the SF to the MI has been studied by some experiments \cite{hung2010slow,bakr2010probing,sherson2010single}. 
In this case, the presence of the Mott shell, formed by a fast equilibration process, prevents the escape of the central excess SF component to the outside, leading to an anomalously 
long time scale of the global equilibration \cite{bernier2011slow,natu2011local,bernier2012slow}. 
For the discussion on a quench from the MI to the SF, the previous theoretical studies \cite{horiguchi2009non,shimizu2018dynamics} did not take into account an effect of a harmonic trap. 
On the other hands, the experiment by Chen \textit{et al.} \cite{chen2011quantum} demonstrated the quantum quench 
in the presence of a harmonic trap, so that considering the inhomogeneous effect is necessary to understand precisely the experimental observations. 

Here, we apply the theory of inhomogeneous Kibble-Zurek mechanism (IKZM) proposed by Ref.~\cite{del2011inhomogeneous} to 
study the quench dynamics of the MI-SF transition in two-dimensional (2D) lattice system with a harmonic confinement. 
A recent work shows that, even for the inhomogeneous system, the universality 
of the quench dynamics can be seen in the quantum dynamics of the one-dimensional Ising model \cite{gomez2019universal}. 
We find that the presence of the Mott lobe in the phase diagram of the BHM provides new features of the IKZM, 
where the ``local" quench time has a nontrivial dependance of the radial coordinate. 
Employing the time-dependent Gutzwiller methods \cite{horiguchi2009non,shimizu2018dynamics}, we simulate the quench dynamics from the MI to the SF and formation of quantized vortices in the superfluid order parameter. 
We find that the quench dynamics is strongly dependent on the frequency of the harmonic trap, where the IKZM is valid for the system in a shallow harmonic trap, while the growth of the SF component is rather adiabatic for a steep harmonic trap. 
When the initial MI has a wedding cake structure, long-lived extra vortices are generated at the interface between the MI domain with different filling factors. This implies the difficulty in experiments to extract the vortices created purely through the KZM.

The paper is organized as follows. Section \ref{KZM} describes a brief review of the KZM for both homogeneous and inhomogenous situations. 
In Sec.~\ref{BHMM}, we introduce the BHM and apply the IKZM to describe the quantum quench from a MI to a SF phase in 
a harmonic trap potential. In Sec.~\ref{gw}, we demonstrate the quench dynamics by using the time-dependent Gutzwiller mean field equation and verify the prediction 
of the IKZM. Section~\ref{concle} devotes to the conclusion. 

\section{Summary of inhomogeneous Kibble-Zurek mechanism}\label{KZM}
The KZM is a theory describing the physical picture of nonequilibrium dynamics and topological defect formations in the system undergoing 
a rapid continous phase transition. 
Here, we briefly review the Kibble-Zurek (KZ) theory and its extended version to inhomogeneous systems introduced by Del Campo \textit{et al.} \cite{del2011inhomogeneous}. 

\subsection{KZM in homogeneous systems}
We suppose that the phase transition is driven by the time-dependent controllable parameter $T(t) = T_c (1 + t/\tau_Q)$, where 
$T_c$ represents the critical point and $\tau_Q$ provides a time scale of a quench. 
Here, we assume that the symmetry-preserved (symmetry-broken) phase exists at $T<T_c$ ($T>T_c$) and 
the system passes the critical point at $t=0$ through the linear ramp of $T(t)$. 
In the vicinity of the transition point, the relaxation time $\tau$ and the correlation 
length $\xi$ in equilibrium diverge as 
\begin{equation}
\tau(\epsilon) = \frac{\tau_0}{|\epsilon|^{z\nu}}, \quad \xi(\epsilon) = \frac{\xi_0}{|\epsilon|^{\nu}}, 
\end{equation} 
where $\tau_0$ and $\xi_0$ are typical scales of time and length, respectively, and $\nu$ and $z$ the critical exponents of a system. 
The time-dependent dimensionless parameter $\epsilon = [T(t) - T_c]/T_c = t/\tau_Q$ describes the deviation from the critical point. 

The dynamics of the phase transition is effectively divided into adiabatic and nonadiabatic regimes, by comparing the velocity with which 
the correlation length would have to increase to maintain its equilibrium value $v_{\xi} = \dot{\xi} = (d\xi/d\epsilon) \dot{\epsilon}$ with 
the propagation speed $s = \xi/\tau$ of the fluctuation. Under the condition $v_{\xi} = s$ at $t=\hat{t}$,
the dynamics can be regarded as adiabatic (nonadiabatic) for $t > |\hat{t}|$ ($t < |\hat{t}|$). 
The time $\hat{t}$ is given by $\hat{t} = (\tau_0 \tau_Q^{z\nu})^{1/(1+z\nu)}$. 
During the time interval $-\hat{t} \leq t \leq \hat{t}$, referred to as a ``frozen region'', the ordered phase develops heterogeneously in the space, 
and the mismatch of the phases of the order parameters leaves the phase defects. 
The size of the generated domains of the ordered phase can be estimated as $\hat{\xi} = \xi(\epsilon(\hat{t}))$ and 
the defect density as $n_\text{def} \sim \hat{\xi}^{d-D}$, where $D$ and $d$ are the dimensions of the space and defect, respectively. 
Thus, we have the power-law relation for $\hat{\xi}$ and $n_\text{def}$ with respect to the quench rate $\tau_Q$ as
\begin{align}
\hat{\xi} &= \xi_0 \left( \frac{\tau_Q}{\tau_0} \right)^{\nu/(1+z\nu)}, \\
 n_\text{def} &= \frac{1}{\xi_0^{D-d}} \left(  \frac{\tau_Q}{\tau_0} \right)^{(D-d) \nu / (1+z\nu)}
\end{align}

\subsection{KZM in inhomogeneous systems (IKZM)}\label{disIKZM}
The above KZ theory can be extended into the inhomogeneous system \cite{del2011inhomogeneous}. 
Keeping in mind the subsequent discussion in which we consider an isotropic harmonic potential, 
we take into account the inhomogeneity through the radial dependence of the critical point as $T_c \to T_c (r)$. 
The dimensionless parameter $\epsilon$ also has a radial dependence 
\begin{equation}
\epsilon(t,r) = \frac{T(t) - T_c(r)}{T_c(r)}.   \label{epsipart1}
\end{equation}
Suppose that the time-dependent parameter $T(t)$ passes the critical point at $t=t_F$. 
Then, the time $t_F$ should be determined locally as $t_F = t_F(r)$; 
the condition $\epsilon(t_F(r),r) = 0$ gives the transition point at the radial position $r$. 
When the controllable parameter changes as $T(t) = T_c(0) (1 + t/\tau_Q)$, predetermined by $T_c(0)$ at $r=0$, we can define the 
\textit{local} quench time as 
\begin{equation}
\tau_Q (r) = \frac{T_c(r)}{T_c(0)} \tau_Q,  \label{localquenchtimeef}
\end{equation}
and get the relation $t_F(r) =  \tau_Q(r) -  \tau_Q$.
By using $\tau_Q(r)$, Eq.~\eqref{epsipart1} can be rewritten as 
\begin{equation}
\epsilon(t,r) = \frac{t+\tau_Q-t_Q(r)}{\tau_Q(r)} =\frac{t-t_F(r)}{\tau_Q(r)}.
\end{equation}
As a result, $\hat{\tau}=\tau(\epsilon(\hat{t}))$ and $\hat{\xi}=\xi(\epsilon(\hat{t}))$ in the KZM also have an radial dependence 
through the replacement of the local quench time $\tau_Q \to \tau_Q(r)$. 

To determine the region in which the KZM may take place in a inhomogeneous system, 
we introduce another characteristic velocity, namely the propagation velocity of the region passing the local transition point. 
This can be estimated as 
\begin{equation}
v_F =  \left| \frac{d \tau_Q(r)}{dr} \right|^{-1} = \frac{T_c(0)}{\tau_Q} \left| \frac{dT_c(r)}{dr} \right|^{-1} . \label{frontvelocity}
\end{equation}
In a homogeneous system, $v_F$ should be infinity. 
Then, the adiabaticity condition can be gained by comparing the propagation velocity $v_F$ and the sound velocity $s$ at $t=\hat{t}$, 
where $s(\hat{t}) \equiv \hat{s}$ is written by 
\begin{equation}
\hat{s} = \frac{\hat{\xi}}{\hat{\tau}} = \frac{\hat{\xi}}{|\hat{t}|} = \frac{\xi_0}{\tau_0} \left| \frac{\tau_0}{\tau_Q(r)} \right|^{\nu (z-1)/(1+\nu z)} .  \label{vhathat}
\end{equation}
The inequality $v_F > \hat{s}$ gives the condition in which the conventional KZM is expected, determining the spatial region 
for the appearance of defects through the KZM. 
Using a particular model, e.g. the BHM as shown below, we shall calculate the threshold value of the radius $r$ from the 
above condition.

\section{The BHM and inhomogeneous quench} \label{BHMM}
In this section, we introduce the BHM to describe the cold bosons in an optical lattice and apply the IKZM to study 
quench dynamics from the MI to the SF in an inhomogeneous situation. 
The inhomogeneity is included by the radial harmonic potential, which exists in typical experimental setups \cite{chen2011quantum,braun2015emergence}. 

\subsection{BHM}
We start from the 2D Bose-Hubbard hamiltonian
\begin{align}
\hat{H} = - J \sum_{\langle i,j \rangle} \left( \hat{b}_j^{\dagger} \hat{b}_i + \text{H.c.} \right) - \sum_j \mu(r) \hat{n}_j \nonumber \\
+ \frac{U}{2} \sum_{j} \hat{n}_j (\hat{n}_j-1), \label{BHMhar}
\end{align}
where $J( > 0)$ represents a tunneling term, $\hat{b}_j$ and $\hat{b}_j^{\dagger}$ the annihilation and creation operators which obey the commutation relation $[\hat{b}_i, \hat{b}_j^{\dagger}] = \delta_{ij}$, $\hat{n}_j =  \hat{b}_j^{\dagger} \hat{b}_j$ the number operator of bosons at a lattice site $j = (j_x,j_y)$, and $U$ the strength of the on-site repulsion between two bosons. The sum of the first term is taken for nearest neighbor sites $\langle i,j \rangle$. 
Since we consider the inhomogeneous 2D system by introducing a harmonic potential, the chemical potential $\mu$ has a radial dependance as
\begin{equation}
 \mu (r) = \mu(0) - \frac{1}{2} k r^2 
 \end{equation}
with the spring constant $k$ and the chemical potential $\mu(0)$ at the origin. 
The radial coordinate $r$ can be represented as $r = a_0 |j|$ with the lattice constant $a_0$. 
The particles are confined within the Thomas-Fermi (TF) radius given by $R_\text{TF} = \sqrt{ 2 \mu(0) / k}$.

\begin{figure}[ht]
\centering
\includegraphics[width=1.0\linewidth]{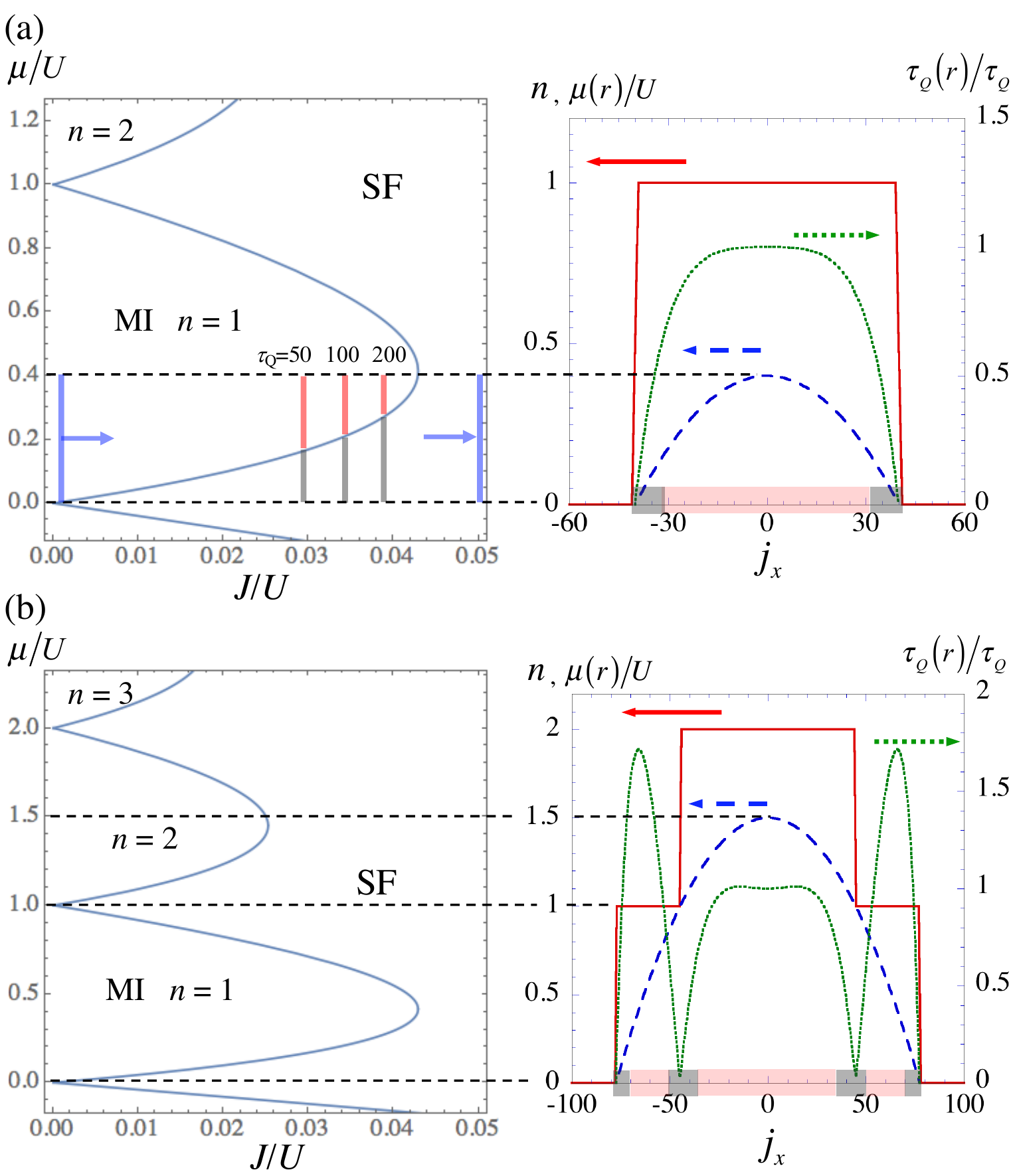}
\caption{ Inhomogeneous quench from the MI to the SF in a harmonically trapped system. The left panels show the ground-state phase diagram of homogeneous BHM for the 2D system in the $J/U$-$\mu/U$ plane, where the phase boundary is given by Eq.~\eqref{criticaltc}. The right panels show the cross section of the equilibrium density profile $n$ (red solid line), the local chemical potential $\mu(r)/U$ (blue dashed curve), and the local quench time $\tau_Q(r)/\tau_Q = T_c(r)/T_c(0)$ (green dotted curve) along the $x$-axis in the presence of the harmonic trap. Here, we set the hopping $J/U=0.001$, the dimensionless spring constant $\tilde{k} = a_0^2 k / U  = 0.0005$, and the central chemical potential (a) $\mu(0)/U=0.4$ and (b) $\mu(0)/U=1.5$. 
In the left panel of (a), the leftmost vertical line at $J/U=0.001$ represents the range of the chemical potential in the trapped system, being rapidly changed to $J/U=0.05$ in the simulations of Sec.~\ref{gw}. The other vertical lines show the range where the transition from the MI to the SF takes place diabatically $(v_F > \hat{s})$ and adiabatically $(v_F <\hat{s})$, represented by red and grey colors, respectively, for the quench time $\tau_Q/\tau_0=50$, $100$, $200$. 
In the right panels, the corresponding diabatic and adiabatic region is shown by the colored bands with red and grey colors, respectively, for $\tau_Q/\tau_0 = 50$ (see the analysis in Sec~\ref{IKZMBHM} and Fig.~\ref{criticalrad}). 
The SF region exists at the very narrow region ($\sim$ 1-site interval) between the domains of the MI phases with different fillings. The SF phase starts to grow during the quench dynamics in the simulations of Sec.~\ref{gw}. }
\label{ikmcpc}
\end{figure}

The ground state of the BHM in a homogeneous system ($k=0$) has been well known, as seen in the standard textbook \cite{lewenstein2012ultracold}. 
There are two ground state phases, namely, the SF and the MI. 
According to the second-order perturbative mean-field theory, the phase boundary can be given by 
\begin{equation}
\frac{J_c}{U} = - \frac{n(n-1) - (\mu/U)(2n-1) + (\mu/U)^2}{Z(1+\mu/U)}.   \label{criticaltc}
\end{equation}
where $n$ is the mean occupation number at each site and $Z$ is the number of nearest neighbors; $Z=4$ in a 2D square lattice system. 
The phase boundary is plotted as shown in Fig.~\ref{ikmcpc}. 
The phase diagram constitutes a well known Mott-lobe structure. 
Although the boundary has been determined more precisely by the Monte-Carlo simulations \cite{capogrosso2008monte}, 
a use of Eq.~\eqref{criticaltc} is enough for our purpose. 
The phase transition across the boundary is known as the second-order transition, and the critical 
exponent belongs to the universality class of the ($d+1$)-dimensional $XY$ model \cite{sachdev2007quantum}. 
In this work, we take $\nu=1/2$ and $z=2$ according to the mean-field theory; the more precise analysis shows that 
the dynamical exponent is $z=1$ on the multicritical point at the tip of the Mott lobe. 

\subsection{IKZM for the BHM} \label{IKZMBHM}
Now, we consider the KZM for the BHM including a harmonic potential. 
Here, the transition from the MI to the SF is caused by a rapid increase of the parameter $J/U$. 
We assume that the on-site interaction $U$ is constant in the followings. 
In the previous literatures, Shimizu \textit{et al.} considered the KZM of the BHM in homogenous system by means of the time-dependent Gutzwiller simulations \cite{shimizu2018dynamics}. 
In the homogeneous situation, however, the unitary time evolution is free from the value of the chemical potential, 
which allows the transition from the MI to the SF only at the tip of the Mott lobe. 
For inhomogeneous cases, the transition point is dependent on the position, since the local chemical potential decreases from the center toward 
the outer region as shown in the right panels of Fig.~\ref{ikmcpc}. 
Also, $\mu(0)$ can be chosen arbitrary in finite size systems, being determined by the particle number in a confining potential. 

In the right panels of Fig.~\ref{ikmcpc}, we plot the density profile for $J/U=0.001$ and (a) $\mu(0) =0.4$ and (b) $\mu(0) =1.5$, where the phase at the center is in a deep MI with $n=1$ and 2, respectively. 
The equilibrium state is calculated by the Gutzwiller ansatz of the wave function, introduced in the next section. 
There are extremely narrow regions of the SF phase between the MI domains with different filling numbers. 
A sudden change of $J/U$ induces the growth of the SF region, which starts from these narrow SF regions. 
This can be seen from the local quench time $\tau_Q(r)$, which is suppressed at the boundary between the MI domains as seen in the right panel of Fig.~\ref{ikmcpc}. 
However, the region in which the KZM takes place is not trivial when we compare the characteristic velocity $v_F$ and $s$, 
because the fluctuations can catch up with the equilibrium values (the dynamics is adiabatic) when $v_F < \hat{s}$, even for small $\tau_Q$. 

\begin{figure}[ht]
\centering
\includegraphics[width=1.0\linewidth]{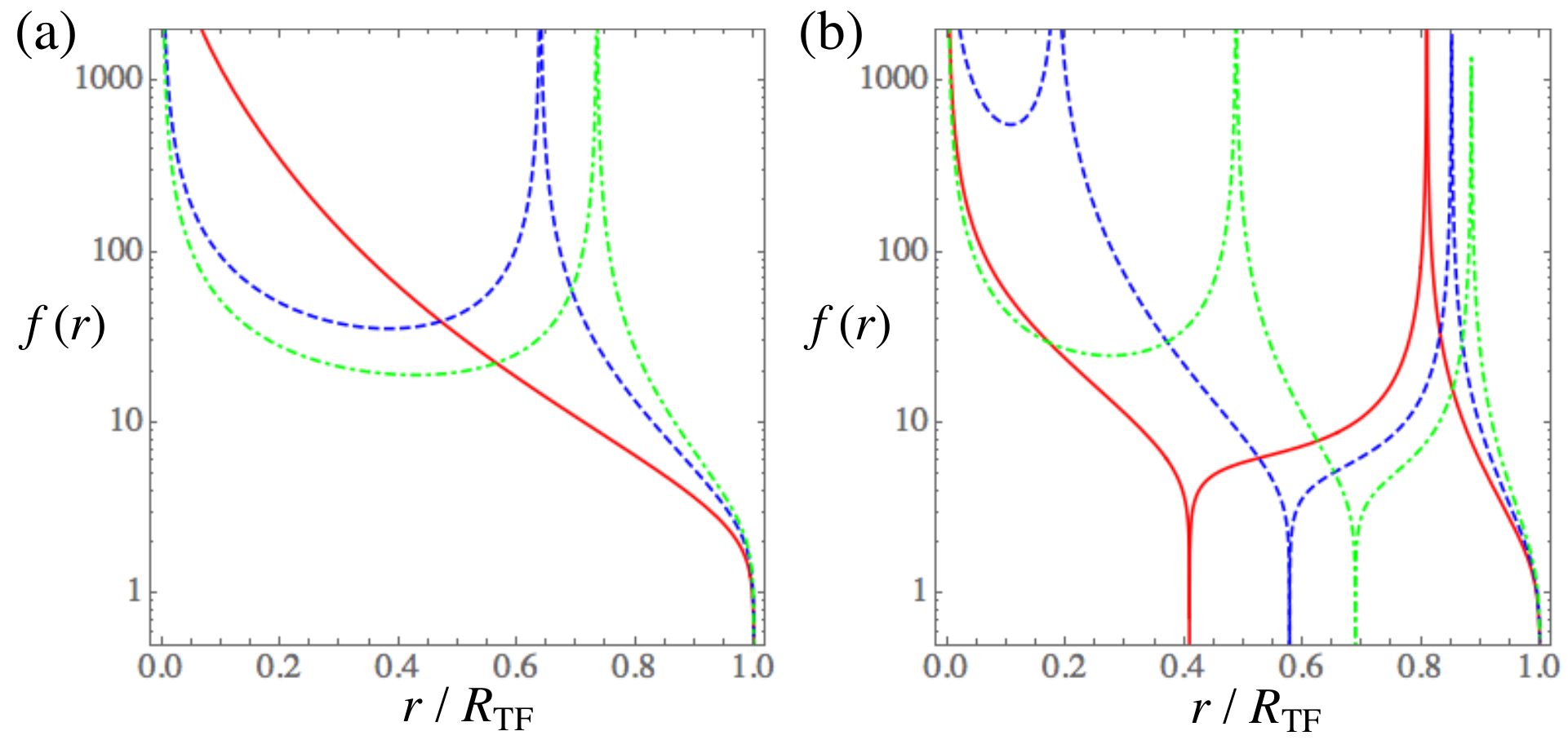}
\caption{(Color online) The function $f(r)$, which is the radial dependence of the ratio $v_F/\hat{v}$, is shown for several values of $\mu(0)$ as a function of $r/R_\text{TF}$. 
The panel (a) corresponds to the transition across the boundary of the $n=1$ Mott-lobe, while (b) across the boundaries of both the $n=1$ and $n=2$ Mott-lobes. 
In (a), the curves represent $\mu(0) = 0.4$ (solid, red), $\mu(0) = 0.7$ (dashed, blue), $\mu(0) = 0.9$ (dotted-dashed, green). 
In (b), the curves represent $\mu(0) = 1.2$ (solid, red), $\mu(0) = 1.5$ (dashed, blue), $\mu(0) = 1.9$ (dotted, green).
}
\label{ikmffunc}
\end{figure}

Let us apply the theory of the IKZM in Sec.~\ref{disIKZM} to the Bose-Hubbard system. 
Our controllable parameter is now $J$ in Eq.~\eqref{BHMhar}.
From Eq.~\eqref{criticaltc}, the critical point $J_c$ has a radial dependence through the chemical potential $J_c(r) = J_c[\mu(r)]$. 
According to Eq.~\eqref{frontvelocity}, we obtain the velocity of the transition front as 
\begin{align}
v_F & = \frac{J_c(0)}{\tau_Q} \left| \frac{dJ_c}{d \mu} \frac{d \mu}{dr} \right|^{-1} \nonumber  \\ 
& = \frac{J_c(0)}{\tau_Q} \left| \frac{-n^2 - n + 1 + 2 (\mu/U) + (\mu/U)^2}{Z(1+\mu/U)^2} k r \right|^{-1}.  \label{vfpropvelo}
\end{align} 
Using Eqs.~\eqref{localquenchtimeef}, \eqref{vhathat} and \eqref{criticaltc}, we can also obtain the expression $\hat{s}$ as a function of $r$. 
The comparison of $v_F$ and $\hat{s}$ determines the condition for the defect formation through the KZM, 
the ratio being represented as 
\begin{equation}
\frac{v_F}{\hat{s}} = A f(r) 
\end{equation}
with the constant 
\begin{equation}
A = \frac{U}{2 \mu(0)} \frac{R_\text{TF}}{\xi_0} \left(  \frac{J_c(0)}{U}  \frac{\tau_0}{\tau_Q} \right)^{(1+\nu)/(1+\nu z)} ,
\end{equation}
determined by the quench time $\tau_Q$ and the chemical potential $\mu(0)$ at $r=0$. The radial dependence is given by 
\begin{equation}
f(r) =  \left( \frac{J_c(r)}{U} \right)^{\frac{\nu (z-1)}{1+\nu z}}  \left| \frac{Z(1+\mu/U)^2 (R_\text{TF}/r)}{-n^2 - n + 1 + 2 (\mu/U) + (\mu/U)^2} \right|  ;
\end{equation}
the function $f(r)$ is shown in Fig.~\ref{ikmffunc} for several values of $\mu(0)$. 


The condition $v_F > \hat{s}$ gives the region in which the vortex formation via the KZM can take place; hereafter we refer to this region as ``KZ region". 
This condition can be obtained by drawing the horizontal line at $A^{-1}$ in Fig.~\ref{ikmffunc} and reading the crossing points given by $f(r) = A^{-1}$, which gives 
the threshold value of the radius $\tilde{r}_c = r_c/ R_\text{TF}$. 
In the calculation, we give the scaled spring constant $\tilde{k} = a_0^2 k / U = 0.0005$ or 0.0025, and assume $\xi_0 \sim a_0$. 
We also confine ourselves to the exponents for the mean-field theory $\nu=1/2$ and $z=2$.

Let us first consider the phase transition across the boundary of the Mott-lobe with $n=1$, corresponding to $\mu(0) < U$ and Fig.~\ref{ikmffunc}(a). 
For $\mu(0) < 0.41 U \equiv \mu_1 $, corresponding to a tip of the Mott-lobe at $r=0$, $f(r)$ decreases monotonically with $r$ and diverges at the origin. 
This divergence is due to the existence of a maximum of $\mu(r)$ at the origin, which causes $d\mu/dr|_{r=0}=0$ and the infinite velocity $v_F$ of the transition front [see Eq.~\eqref{vfpropvelo}]. 
For $\mu(0) > \mu_1$ there are two divergent peaks; one is at the origin and the other corresponds to the tip of 
the Mott lobe, at which $J_c(r)$ becomes a maximum with respect to $\mu(r)$ and thus $dJ_c/d\mu = 0$. 
This nontrivial radial dependence of the front velocity $v_F$ is a characteristic feature of the Bose-Hubbard system. 

\begin{figure*}[ht]
\centering
\includegraphics[width=0.8\linewidth]{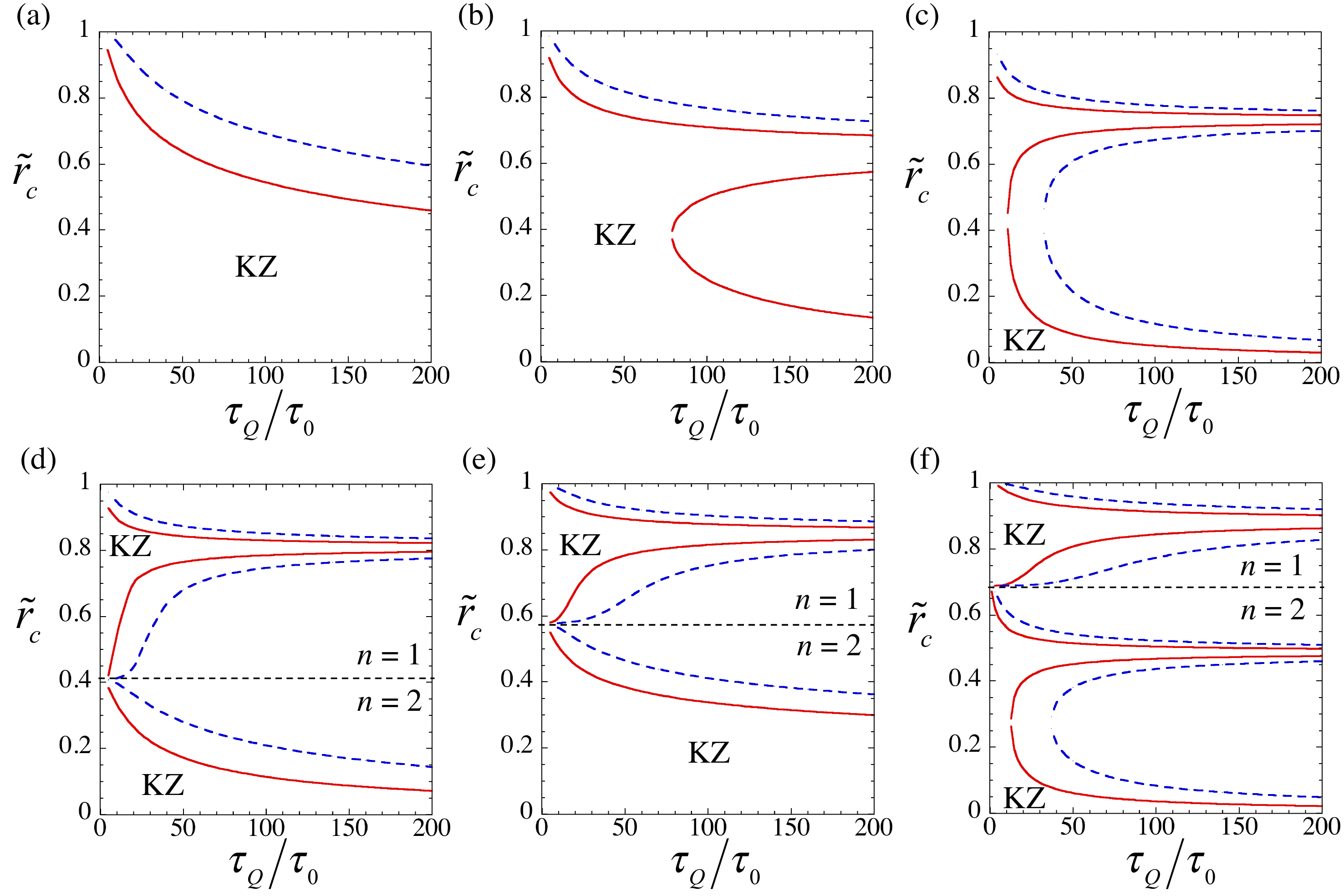}
\caption{The critical radius giving the KZ region in a harmonically trapped system with respect to the quench time $\tau_Q/\tau_0$. We show the results for two scaled spring constants $\tilde{k} = a_0^2 k / U = 0.0025$ and $0.0005$ by red-solid curves and blue-dashed curves, respectively. The chemical potentials at the center are (a) $\mu(0)=0.4$, (b) $\mu(0)=0.7$, (c) $\mu(0)=0.9$, (d) $\mu(0)=1.2$, (e) $\mu(0)=1.5$, (f) $\mu(0)=1.9$.
}
\label{criticalrad}
\end{figure*}
Figure~\ref{criticalrad} shows the threshold radius as a function of $\tau_Q$, obtained by taking the crossing points of the $f(r)$-curve and the $A^{-1}$-line in Fig.~\ref{ikmffunc}. 
Here, we plot the results for the different values of $\tilde{k} =0.0025$ and $\tilde{k} =0.0005$.
The KZ region is strongly dependent on the value of $\mu(0)$. For $\mu(0) < \mu_1$ the critical radius decreases monotonically with increasing $\tau_Q$, namely the KZ region gets narrow from outside [Fig.~\ref{criticalrad}(a)]. 
We show the KZ region for $\mu(0) =0.4$, $\tilde{k} =0.0005$ and $\tau_Q /\tau_0 = 50$ by the shaded band in the right panel of Fig.~\ref{ikmcpc}(a), where $\tilde{r}_c \simeq 0.8$. 
For $\mu(0)>\mu_1$, however, there appears another non-KZ region around the intermediate region at $r \sim 0.4 R_\text{TF}$ [Fig.~\ref{criticalrad}(b)] for $\tau_Q/\tau_0 > 100$. This means that the KZM occurs at the central region and the ring-shaped region separated from the central KZ region for a relatively slow quench. 
Near $\mu(0) \lesssim U$ the KZ regions are further shrunk to the narrow region around the center and $r \sim 0.75R_\text{TF}$ as shown in Fig.~\ref{criticalrad}(c). 
Thus, the IKZM would predict a very different behavior of the KZ dynamics from MI to SF depending on whether the central chemical potential $\mu(0)$ is smaller or larger than the $\mu_1$.

When the transition includes the several Mott-lobes, the situation becomes more complicated, as shown in Fig.~\ref{ikmffunc}(b) and Fig.~\ref{criticalrad}(d)-(f) for the case including both the $n=1$ and $n=2$ Mott-lobes. 
The KZ region can be obtained similarly by considering that the velocity $v_F$ diverges both at the center and at the tips of the Mott-lobes. 
The KZ region for the $n=2$ MI domain is maximized when $\mu(0)$ is located near the tip of the Mott lobe, while it takes place only near the tip of the Mott lobe for the surrounding $n=1$ MI domain. We show the KZ region for $\mu(0) = 1.9$ [Fig.~\ref{criticalrad}(f)] and $\tau_Q/\tau_0=50$ by the shaded region in the right panel of Fig.~\ref{ikmcpc}(b), where the KZ region corresponds to $0 \leq r/R_\text{TF} \leq 0.16$, $0.38 \leq r/R_\text{TF} \leq 0.54$ and $0.71 \leq r/R_\text{TF} \leq 0.96$.

\section{Time-dependent Gutzwiller analysis of quench dynamics}\label{gw}
To demonstrate the above prediction of the IKZM, we make numerical simulations of the quench dynamics described by the BHM with the harmonic potential of Eq.~\eqref{BHMhar}. 
To study the real time evolution in the 2D system, we employ the time-dependent Gutzwiller method. 
The Gutzwiller approximation is based on the assumption of a variational state which has a product form in which correlations between different sites factorize into single-site state vectors, which does not capture correlations involving different sites. 
Nevertheless, because of its simplicity, it is useful method for gaining an understanding of physics for regimes where exact numerical results are not easily obtained, e.g., out-of-equilibrium dynamics in dimensions higher than unity \cite{natu2011local,zakrzewski2005mean,snoek2007two,lundh2011mott,krutitsky2011excitation,snoek2012collective,rapp2013mean,yan2017equilibration}.
Also some studies have used this method to study the quench dynamics relevant to the KZM \cite{horiguchi2009non,shimizu2018dynamics,shimizu2018dynamics2,zhou2020quench}. 
In this work, we employ this method to consider quench dynamics in the 2D system, thus giving qualitative discussion on how the inhomogeneous phase transition occurs and whether the IKZM can apply the simulation results or not. Within the mean-field approximation, only local quantum fluctuation is taken into account, whereas long-wavelength fluctuations are important in a lower dimension, because they destroy true superfluid long-range order. For the trapped system at a zero temperature, however, the concern is less important since the harmonic potential provides a natural cutoff for the long-wavelength fluctuation.

The Gutzwiller ansatz for the many-body wave function is written as
\begin{eqnarray}
\ket{\Psi_\text{G}(t)}=\prod_{j}\sum_{n}f_{j,n}(t)\ket{n}_{j}, 
\end{eqnarray}
where $f_{j,n}(t)$ represents the complex coefficients for the number state $| n \rangle_{j}$ at the $j$-th site. 
The ansatz corresponds to a mean-field approximation by ignoring the correlation between different sites, 
Under the variational principle, we minimize $\bra{\Psi_{G}}\hat{H}-i \hbar \frac{d}{dt}\ket{\Psi_{G}}$ with respect to $f^{\ast}_{j,n}$ to obtain the time-dependent Gutzwiller equation.
\begin{eqnarray}
i \hbar \frac{d f_{j,n}}{dt}=-J \sum_{\langle i,j \rangle}\biggl{[}\sqrt{n}\psi_{i}  f_{j,n-1}
+\sqrt{n+1} \psi^{\ast}_{i} f_{j,n+1} \biggl{]} \nonumber \\
+\biggl{[} \mu(r) n+\frac{U}{2}n(n-1)\biggl{]}f_{j,n}.  \label{timedepGW}
\end{eqnarray}
Here, the SF order parameter is given as
\begin{eqnarray}
\psi_{j}=\braket{\hat{b}_{j}}=\sum_{n}\sqrt{n+1}f^{\ast}_{j,n}f_{j,n+1},
\end{eqnarray}
Equation \eqref{timedepGW} is solved numerically by the Crank-Nicholson method, where the Neumann boundary condition is used at the edge of the simulation system. 
We have confirmed that the total energy and the norm is conserved during the time evolution in the case of constant values of $J$ and $U$. 
 
In order to induce the quench from the MI to the SF phase and study the IKZM, the hopping amplitude is varied as 
\begin{equation}
\frac{J(t) - J_c(0)}{J_c(0)} = \frac{t}{\tau_Q}  \label{quenchmethod}
\end{equation} 
with the critical hopping $J_c(0)$ at $r=0$, determined by Eq.~\eqref{criticaltc} within the perturbation theory. 
The initial states of the simulations are prepared in the deep MI region, where we choose the hopping amplitude as $J(t=0) = 0.001 U$. 
Then, the value $J(t)$ is linearly increased according to Eq.~\eqref{quenchmethod} to the final value $J(t=t_f)=0.05U$ at which the whole system is in the SF phase. 
The slope of the ramping up of $J(t)$ is given by $J_c(0)/\tau_Q$. 

For the initial Mott state, we set the phases of $\{ f_{j,n} (t=0) \}$ fully random to mimic the quantum fluctuation in the mean-field method \cite{shimizu2018dynamics,shimizu2018dynamics2,zhou2020quench}. 
However, the amplitude of the initial noise plays a crucial role to the growth time of the superfluid order parameter $\psi$. Thus, it is difficult to determine the observation time of the vortex number, called as the KZM time, from the $\psi$-based protocols \cite{shimizu2018dynamics,zhou2020quench}. 
Furthermore, the amplitude of the superfluid density also grows inhomogeneously in our inhomogeneous system, we cannot determine uniquely the KZM time fitted to the whole system. 
Thus, we do not discuss the scaling property of the defect density with respect to the global quench time $\tau_Q$. 
Alternately, we discuss how the vortices nucleate from the inhomogeneous system and the difference from the homogeneous situation.

When $J(t)$ is increased, topological defects (vortices) can be created. 
The vortices are identified by calculating the current density $j_{kl} = - \sqrt{n_k n_l} \sin(\theta_k - \theta_l)$ between $k$ and $l$-sites, where $k(\neq l)$ represents the label of the spatial grids in the 2D space; 
if all $j_{kl}$ along a certain minimal loop $(k, l) \to( k+1,l) \to (k+1,l+1) \to (k,l+1) \to (k,l)$ have the same sign, a vortex exists at the inside of the loop. 
Due to the random phases in the initial state, there are vortices even before the SF order parameter develops. 


In the following, we consider the situations with three different values of the chemical potential $\mu(0)$ at the origin, namely 
$\mu(0)/U=0.4$, $\mu(0)/U=0.9$, $\mu(0)/U=1.5$, and demonstrate the quench dynamics from MI to SF in the harmonic trap potential. 

\subsection{$\mu(0)/U=0.4$}
\begin{figure}[ht]
\centering
\includegraphics[width=1.0\linewidth]{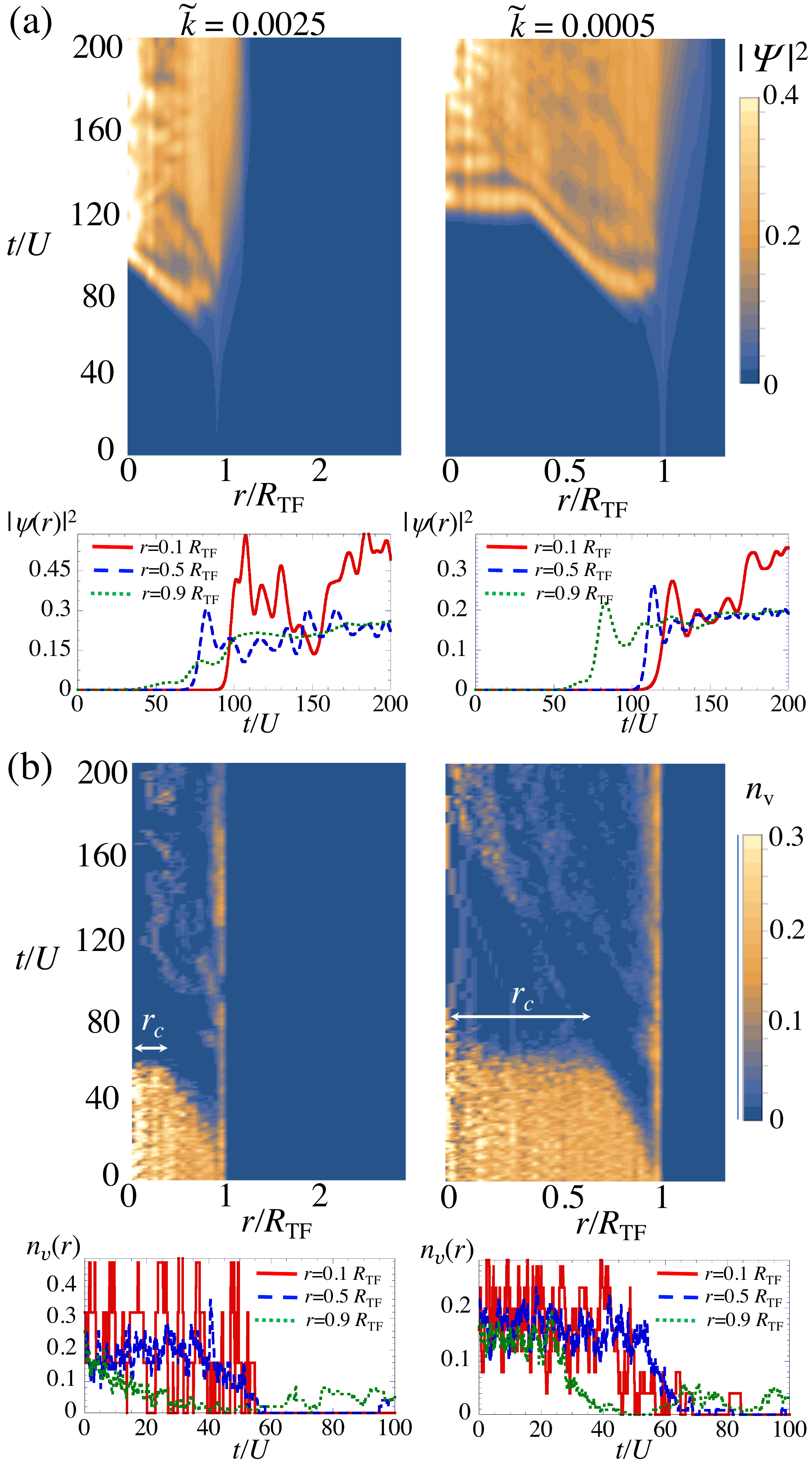}
\caption{The time development of the radial SF density $|\psi(r)|^2$ (a) and the radial vortex density $n_v(r)$ (b) for $\mu(0)/U = 0.4$ and $\tau_Q = 50$. 
The left panels and right panels correspond to $\tilde{k}=0.0025$ and $\tilde{k}=0.0005$, respectively. 
In each figure, we also show the cross section for several values of the radius along the time axis. 
The Thomas-Fermi radius is $R_\text{TF}/a_{0} = 18$ for (a) and $R_\text{TF}/a_{0} = 40$ for (b). }
\label{dmu04}
\end{figure}
In this case, the quench of $J(t)$ induces the quantum phase transition from the periphery of the system. 
Under the local approximation, the timing which passes through the transition point is latest at the center, where the critical value is $J_c(0)/U = 0.0429$.   
To see the inhomogeneity of the transition dynamics, we represent the time development of the radial distribution of the 
superfluid density $|\psi(r)|^2 =(2\pi)^{-1} \int d\theta |\psi(r, \theta)|^2$ and the radial vortex density $n_v(r) =(2\pi)^{-1} \int d\theta n_v(r,\theta)$. 
Here, the vortex density is taken only for vortices within the Thomas-Fermi radius $R_\text{TF}$. 
In Fig.~\ref{dmu04}, we plot the results for $\tilde{k}=0.0025$ and $\tilde{k}=0.0005$ for $\tau_Q = 50$. 
For the steep trap $\tilde{k}=0.0025$, the superfluid component arises from the periphery of the MI domain. 
The superfluid density grows from the periphery to the center, along with the transition line of the Mott-lobe. 
Since the MI has a strong phase fluctuation, the region within the Thomas-Fermi radius is initially filled with vortices. 
After the quench, due to the appearance of the phase coherence in the superfluid phase, 
the vortices are disappeared from the outside to the inside smoothly, as shown in the left panel of Fig.~\ref{dmu04}(b). 
The mechanism of the disappearance of vortices is the vortex--anti-vortex pair annihilation. 

For the shallow trap $\tilde{k}=0.0005$, the dynamics is slightly different. 
The superfluid density begins to grow and the vortices are erased gradually from the outside. 
At a certain time, however, the transition occurs homogeneously within the radius $r_c$ smaller than $R_\text{TF}$. 
The right panel of Fig.~\ref{dmu04}(b) clearly indicates that the most of vortices due to the random noise in the MI are suddenly disappeared at $t/U=60$ within the radius $r_c$. 
From the development of the vortex density, we can identify the region of the non-adiabatic KZM. 
Thus, the applicability of the IKZM theory is strongly dependent on the shallowness of the harmonic trap potential. 

 \begin{figure}[ht]
\centering
\includegraphics[width=1.0\linewidth]{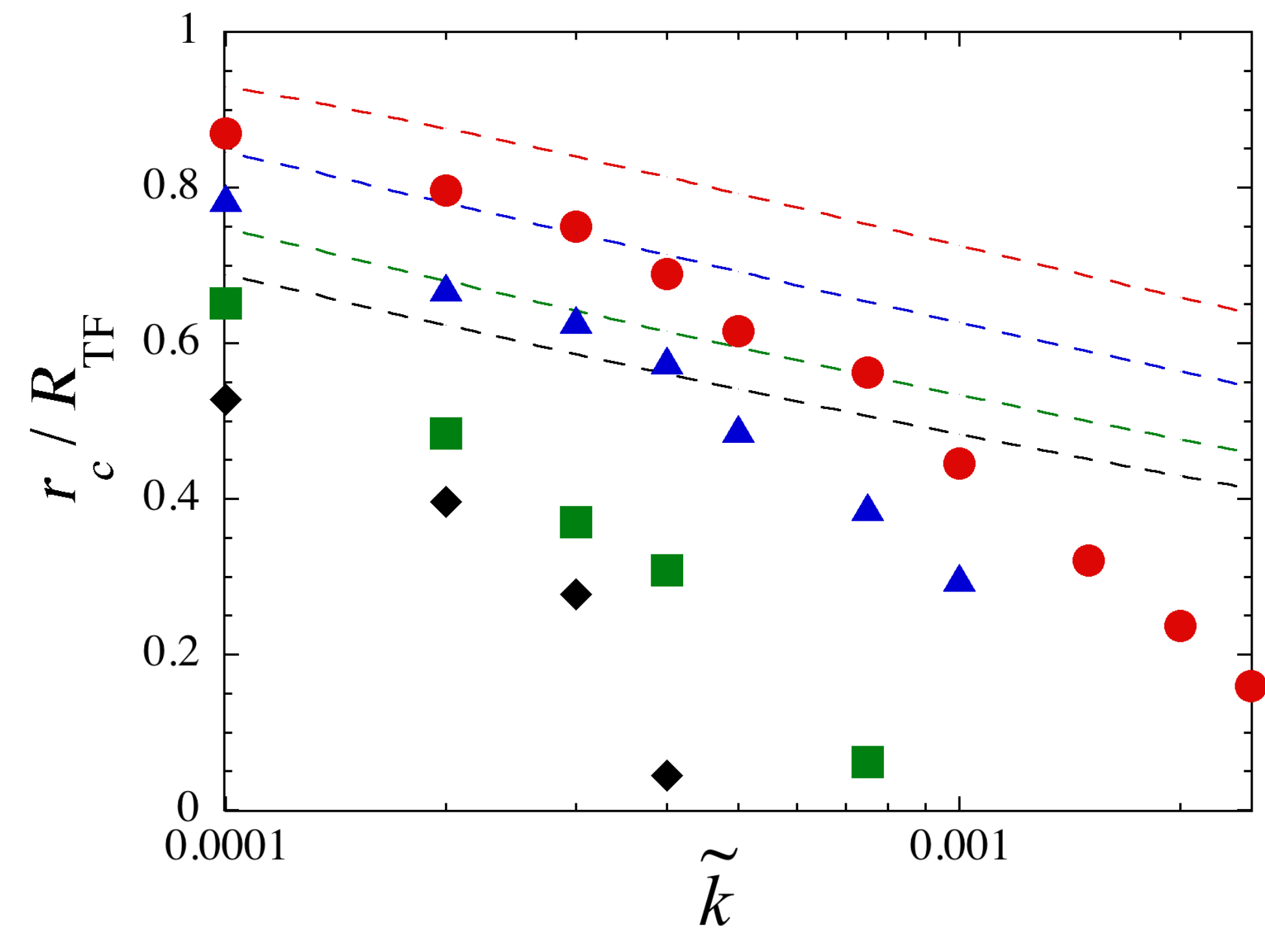}
\caption{The critical radius giving the KZ region as a function of the spring constant $\tilde{k}$ of the harmonic trap. The plots are obtained from the numerical simulations, where the global quench times are given by $\tau_Q=50$ (red circles), $100$ (blue triangles), $200$ (green squares), and $300$ (black diamonds). The dashed lines correspond to $r_c$ given by the IKZM theory, where $\tau_Q$ = 50, 100, 200, 300 from top to bottom. 
}
\label{criticalradGW}
\end{figure}
We find that the critical radius $r_c$ is dependent on the quench time and the trap spring constant $\tilde{k}$. 
Figure \ref{criticalradGW} shows the critical radius $r_c/R_\text{TF}$, extracted from the development of $n_v(r)$, as a function of the spring constant $\tilde{k}$ for several quench times $\tau_Q$. 
The critical radius $r_c/R_\text{TF}$ increases as $\tilde{k}$ decreases, as expected from the fact that the transition approaches to the homogeneous limit $r_c/R_\text{TF} \to 1$ for $\tilde{k} \to 0$. 
The slow quench suppresses the appearance of the KZ region, where $r_c = 0$ means that the transition is adiabatic. 
We also plot $r_c$ obtained from the IKZM in Sec.~\ref{IKZMBHM}. 
Although the numerical result underestimates the KZ region obtained from the theory, it approaches the theoretical estimation for smaller values of $\tilde{k}$, i.e., a shallow trap potential, and the fast quench. 
The difficulty of the application to IKZM in our case might be the breakdown of the local approximation of the chemical potential to evaluate the MI-SF phase diagram. 

As can be seen in the time evolution of $n_v$, the vortices caused by the original phase fluctuation are almost disappeared along with the growth of the superfluid density. The vortices surviving after the growth of the phase coherence can be regarded as the vortices created by the KZM. 
However, we see that the vortices are invaded from outside of the Thomas-Fermi radius even after the superfluid order parameter is sufficiently grown. 
These vortices are not relevant to the KZM.

\subsection{$\mu(0)/U=0.9$}
\begin{figure}[ht]
\centering
\includegraphics[width=1.0\linewidth]{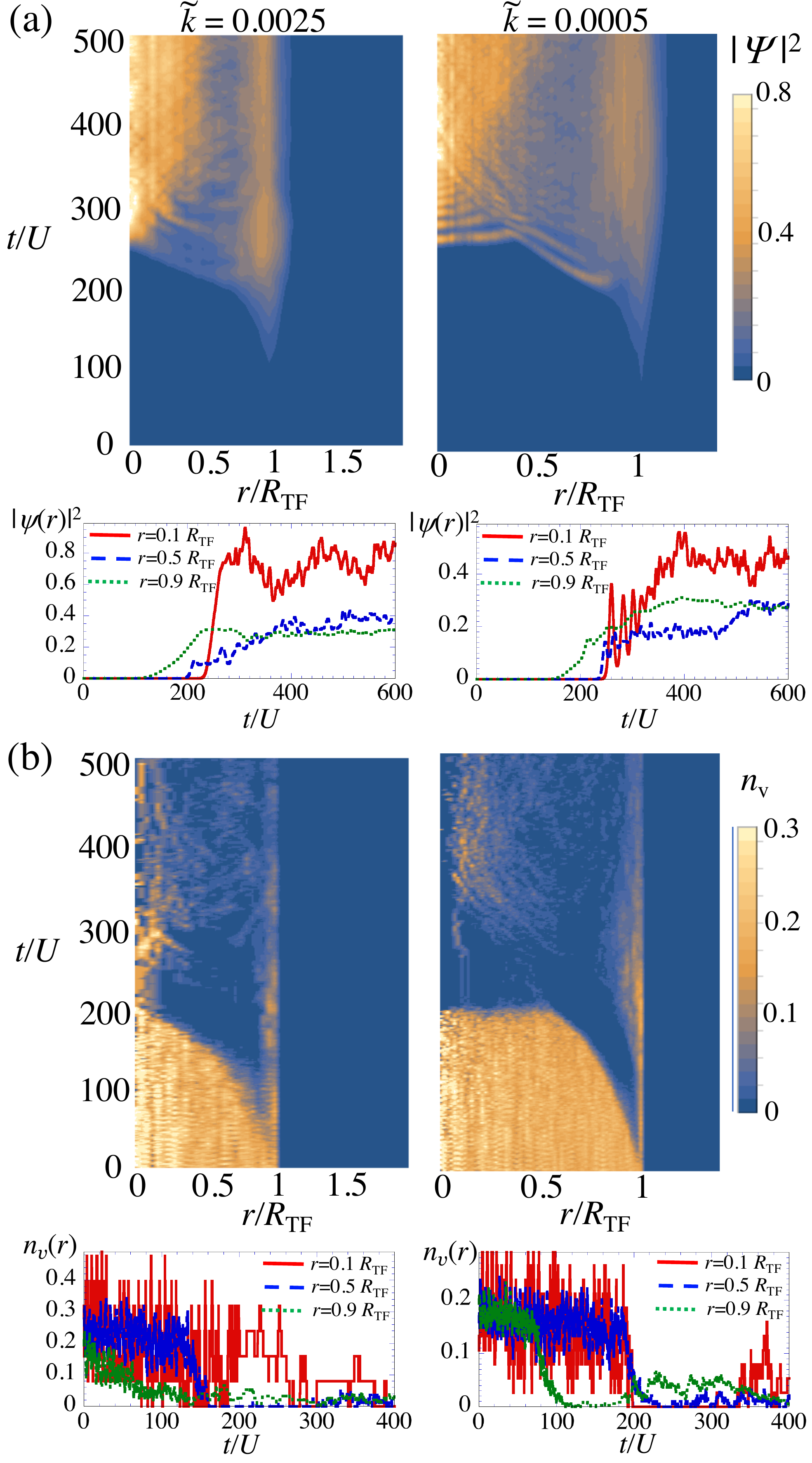}
\caption{The time development of the radial SF density $|\psi(r)|^2$ (a) and the radial vortex density $n_v(r)$ (b) for $\mu(0)/U = 0.9$ and $\tau_Q = 50$. 
The left panels and right panels correspond to $\tilde{k}=0.0025$ and $\tilde{k}=0.0005$, respectively. 
In each figure, we also show the cross section for several values of the radius along the time axis. 
The Thomas-Fermi radius is $R_\text{TF}/a_{0} = 27$ for (a) and $R_\text{TF}/a_{0} = 60$ for (b).}
\label{dmu09}
\end{figure}
Next, we consider the case $\mu(0)/U=0.9$, where the MI-SF transition is expected to start both at the periphery and the center, according to the phase diagram of Fig.~\ref{ikmcpc}. 
Figure \ref{dmu09} shows the similar plot with Fig.~\ref{dmu04} for steep and shallow trap cases and $\tau_Q = 50$. 
The time scale of the dynamics becomes longer than that of Fig.~\ref{dmu04} since, according to Eq.~\eqref{quenchmethod}, the slope of $J(t)$ becomes gentle due to small critical value $J_c(0)/U = 0.0118$ at the center. 
In the both cases of the trap, we obtain the results similar to Fig.~\ref{dmu04}. 
Contrary to expectations, the superfluid component grows only from the outside, not from the center. 
This is due to the breakdown of the local argument. 
Since the Gutzwiller equation conserves the mean total particle number, the phase transition at the central region takes place with the fixed particle density $n  = 1$. 
Then the transition from the $n=1$ MI to the SF is prohibited, where the equilibrium local particle number is determined by the local chemical potential in the SF phase and should be increased more than unity. 
However, the central region is embedded deeply in the MI so that the change of the particle density cannot occur. 
On the other hands, at the boundary of $n=1$ and $n=0$ MI domains $(r \simeq R_\text{TF})$ a small fraction of the SF order parameter exist, which can grow through the quench since there is no gap of the SF order parameter determined by the local chemical potential. 
Therefore, the inhomogeneous nature of the quench dynamics is important only for the transition from the outside in the number-conserving simulations. 

The long-time development shows that there emerges a depression of the superfluid density around $r=0.7R_\text{TF}$, where the chemical potential $\mu(0.7R_\text{TF})$ corresponds to the tip of the Mott lobe. 
This radial position can exhibit an anomalous behavior because the $v_F/\hat{s}$ diverges but the local quench time becomes the longest one. 
We see that, in the later stage of the dynamics, the vortices are accumulated in the ring-shaped region around  $r \sim 0.7R_\text{TF}$. 
These behaviors are not changed for different values of the global quench time $\tau_Q$.

\subsection{$\mu(0)/U=1.5$}
\begin{figure}[ht]
\centering
\includegraphics[width=1.0\linewidth]{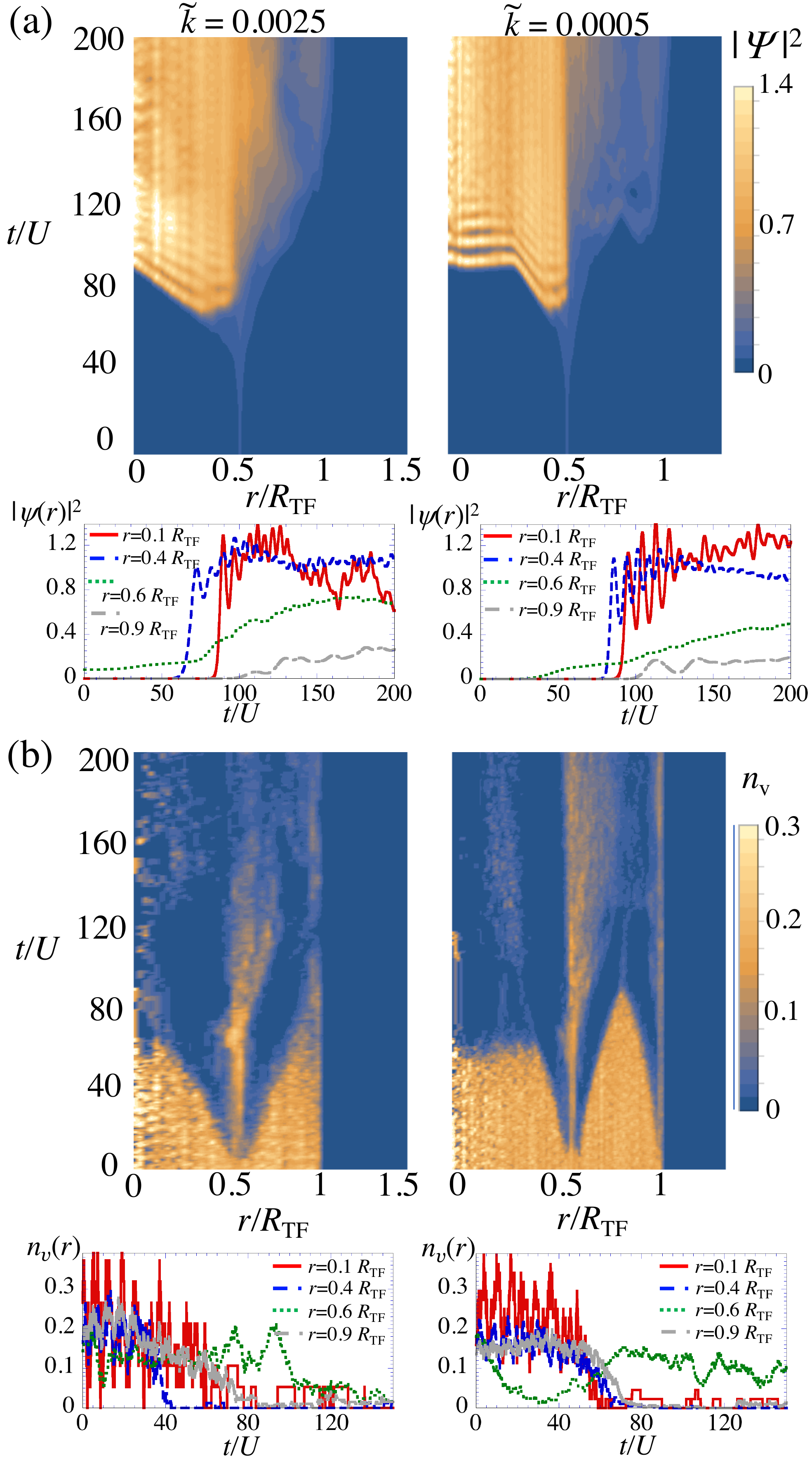}
\caption{The time development of the radial SF density $|\psi(r)|^2$ (a) and the radial vortex density $n_v(r)$ (b) for $\mu(0)/U = 1.5$ and $\tau_Q = 50$. 
The left panels and right panels correspond to $\tilde{k}=0.0025$ and $\tilde{k}=0.0005$, respectively. 
In each figure, we also show the cross section for several values of the radius along the time axis. 
The Thomas-Fermi radius is $R_\text{TF}/a_{0} = 35$ for (a) and $R_\text{TF}/a_{0} = 78$ for (b).
}
\label{dmu15}
\end{figure}
Finally, we study the quench dynamics for $\mu(0)/U = 1.5$, where the initial state consists of a wedding cake of the $n=2$ and $n=1$ MI islands. 
The critical value at the center is given by $J_c(0)/U=0.025$.
As seen in Fig.~\ref{dmu15}(a), for both the shallow and steep trap case, the SF component grows from both the periphery and the boundary between the $n=2$ and $n=1$ MI domains, where the very small fraction of the SF density exists even in the initial state. 
For the steep trap, the SF order grows continuously from these boundary to the inside of the respective MI domain. 
A similar behavior can be seen for the shallow trap case, but a clear separation of the SF density can be seen between the inner and outer regions. 
From Fig.~\ref{dmu15}(b), there are small central regions that are relevant to the KZM, where most of the noisy vortices are disappeared homogeneously. 
In the right panel of Fig.~\ref{dmu15}(b), the transition takes place adiabatically in the surrounding $n=1$ MI domain since the local quench time is longer than the inner region as seen in Fig.~\ref{ikmcpc}.
In addition to these observations, we can see that long-lived vortices are generated from the boundary between the $n=2$ and $n=1$ MI. 
Since the SF order parameters have the different origins in the two regions, the discontinuity of the SF phase may yields the vortices. 
These vortices can have a long life time compared with the vortices arising from the random phase distribution; 
the latter are soon disappeared via pair annihilation. 
This fact provides a difficulty for the experimental identification of vortices purely through the KZM.

\section{Conclusion}\label{concle}
In conclusion, we consider the quench dynamics from the MI to the SF of bosons in an optical lattice and a harmonic trap, which is usually utilized in the cold atom experiments, and the applicability of the IKZM \cite{del2011inhomogeneous}. 
Due to the nontrivial radial dependance of the transition point, we can identify the region where the KZM is expected by applying the theory of the IKZM to the system of the BHM. 
The simulation of the BHM, based on the time-dependent GW equation, demonstrates the rich phenomenology of the quench dynamics, where the IKZM is applicable only for the system with a shallow harmonic trap. 
The inhomogeneity of the system may cause the unexpected generation of the vortices, from the periphery of the Thomas-Fermi radius as well as the phase boundary between $n=1$ and $n=2$ MI domains. 
These bring the difficulty for quantitative evaluation of the KZM in the cold atom system in OL and the harmonic trap potential. 
More precise simulations including precisely quantum fluctuations in the initial state, e.g., through the truncated Wigner approximation, remain for future work. 
In the 1D case, calculations based on the Matrix product state can include the quantum fluctuation reliably, enabling us to study the inhomogeneous nature of the KZ scaling, as reported in \cite{gomez2019universal}, by measuring the number of quantum solitons after the quantum quench.

\acknowledgements
We would like to thank I. Danshita, Y. Kondo, I. Ichinose, and I-K. Liu for comments and fruitful discussions. 
The work of K.K. is supported by KAKENHI from the Japan Society for the Promotion of Science (JSPS) Grant-in- Aid for Scientific Research (KAKENHI Grant No. 18K03472). 

\bibliographystyle{apsrev4}
\let\itshape\upshape
\bibliography{reference}

\providecommand{\noopsort}[1]{}\providecommand{\singleletter}[1]{#1}%
\begin{thebibliography}{42}%
\makeatletter
\providecommand \@ifxundefined [1]{%
 \@ifx{#1\undefined}
}%
\providecommand \@ifnum [1]{%
 \ifnum #1\expandafter \@firstoftwo
 \else \expandafter \@secondoftwo
 \fi
}%
\providecommand \@ifx [1]{%
 \ifx #1\expandafter \@firstoftwo
 \else \expandafter \@secondoftwo
 \fi
}%
\providecommand \natexlab [1]{#1}%
\providecommand \emph  [1]{``#1''}%
\providecommand \bibnamefont  [1]{#1}%
\providecommand \bibfnamefont [1]{#1}%
\providecommand \citenamefont [1]{#1}%
\providecommand \href@noop [0]{\@secondoftwo}%
\providecommand \href [0]{\begingroup \@sanitize@url \@href}%
\providecommand \@href[1]{\@@startlink{#1}\@@href}%
\providecommand \@@href[1]{\endgroup#1\@@endlink}%
\providecommand \@sanitize@url [0]{\catcode `\\12\catcode `\$12\catcode
  `\&12\catcode `\#12\catcode `\^12\catcode `\_12\catcode `\%12\relax}%
\providecommand \@@startlink[1]{}%
\providecommand \@@endlink[0]{}%
\providecommand \url  [0]{\begingroup\@sanitize@url \@url }%
\providecommand \@url [1]{\endgroup\@href {#1}{\urlprefix }}%
\providecommand \urlprefix  [0]{URL }%
\providecommand \Eprint [0]{\href }%
\providecommand \doibase [0]{http://dx.doi.org/}%
\providecommand \selectlanguage [0]{\@gobble}%
\providecommand \bibinfo  [0]{\@secondoftwo}%
\providecommand \bibfield  [0]{\@secondoftwo}%
\providecommand \translation [1]{[#1]}%
\providecommand \BibitemOpen [0]{}%
\providecommand \bibitemStop [0]{}%
\providecommand \bibitemNoStop [0]{.\EOS\space}%
\providecommand \EOS [0]{\spacefactor3000\relax}%
\providecommand \BibitemShut  [1]{\csname bibitem#1\endcsname}%
\let\auto@bib@innerbib\@empty
\bibitem [{\citenamefont {Lewenstein}\ \emph {et~al.}(2012)\citenamefont
  {Lewenstein}, \citenamefont {Sanpera},\ and\ \citenamefont
  {Ahufinger}}]{lewenstein2012ultracold}%
  \BibitemOpen
  \bibfield  {author} {\bibinfo {author} {\bibfnamefont {M.}~\bibnamefont
  {Lewenstein}}, \bibinfo {author} {\bibfnamefont {A.}~\bibnamefont {Sanpera}},
  \ and\ \bibinfo {author} {\bibfnamefont {V.}~\bibnamefont {Ahufinger}},\
  }\href@noop {} {\emph {\bibinfo {title} {Ultracold Atoms in Optical Lattices:
  Simulating quantum many-body systems}}}\ (\bibinfo  {publisher} {Oxford
  University Press},\ \bibinfo {year} {2012})\BibitemShut {NoStop}%
\bibitem [{\citenamefont {Polkovnikov}\ \emph {et~al.}(2011)\citenamefont
  {Polkovnikov}, \citenamefont {Sengupta}, \citenamefont {Silva},\ and\
  \citenamefont {Vengalattore}}]{polkovnikov2011colloquium}%
  \BibitemOpen
  \bibfield  {author} {\bibinfo {author} {\bibfnamefont {A.}~\bibnamefont
  {Polkovnikov}}, \bibinfo {author} {\bibfnamefont {K.}~\bibnamefont
  {Sengupta}}, \bibinfo {author} {\bibfnamefont {A.}~\bibnamefont {Silva}}, \
  and\ \bibinfo {author} {\bibfnamefont {M.}~\bibnamefont {Vengalattore}},\
  }\bibfield  {title} {\emph {\bibinfo {title} {Colloquium: Nonequilibrium
  dynamics of closed interacting quantum systems},}\ }\href@noop {} {\bibfield
  {journal} {\bibinfo  {journal} {Reviews of Modern Physics}\ }\textbf
  {\bibinfo {volume} {83}},\ \bibinfo {pages} {863} (\bibinfo {year}
  {2011})}\BibitemShut {NoStop}%
\bibitem [{\citenamefont {Kennett}(2013)}]{kennett2013out}%
  \BibitemOpen
  \bibfield  {author} {\bibinfo {author} {\bibfnamefont {M.~P.}\ \bibnamefont
  {Kennett}},\ }\bibfield  {title} {\emph {\bibinfo {title} {Out-of-equilibrium
  dynamics of the bose-hubbard model},}\ }\href@noop {} {\bibfield  {journal}
  {\bibinfo  {journal} {ISRN Condensed Matter Physics}\ }\textbf {\bibinfo
  {volume} {2013}} (\bibinfo {year} {2013})}\BibitemShut {NoStop}%
\bibitem [{\citenamefont {Kibble}(1976)}]{kibble1976topology}%
  \BibitemOpen
  \bibfield  {author} {\bibinfo {author} {\bibfnamefont {T.~W.}\ \bibnamefont
  {Kibble}},\ }\bibfield  {title} {\emph {\bibinfo {title} {Topology of cosmic
  domains and strings},}\ }\href@noop {} {\bibfield  {journal} {\bibinfo
  {journal} {Journal of Physics A: Mathematical and General}\ }\textbf
  {\bibinfo {volume} {9}},\ \bibinfo {pages} {1387} (\bibinfo {year}
  {1976})}\BibitemShut {NoStop}%
\bibitem [{\citenamefont {Zurek}(1985)}]{zurek1985cosmological}%
  \BibitemOpen
  \bibfield  {author} {\bibinfo {author} {\bibfnamefont {W.~H.}\ \bibnamefont
  {Zurek}},\ }\bibfield  {title} {\emph {\bibinfo {title} {Cosmological
  experiments in superfluid helium?}}\ }\href@noop {} {\bibfield  {journal}
  {\bibinfo  {journal} {Nature}\ }\textbf {\bibinfo {volume} {317}},\ \bibinfo
  {pages} {505} (\bibinfo {year} {1985})}\BibitemShut {NoStop}%
\bibitem [{\citenamefont {Zurek}(1996)}]{zurek1996cosmological}%
  \BibitemOpen
  \bibfield  {author} {\bibinfo {author} {\bibfnamefont {W.~H.}\ \bibnamefont
  {Zurek}},\ }\bibfield  {title} {\emph {\bibinfo {title} {Cosmological
  experiments in condensed matter systems},}\ }\href@noop {} {\bibfield
  {journal} {\bibinfo  {journal} {Physics Reports}\ }\textbf {\bibinfo {volume}
  {276}},\ \bibinfo {pages} {177} (\bibinfo {year} {1996})}\BibitemShut
  {NoStop}%
\bibitem [{\citenamefont {CAMPO}\ and\ \citenamefont
  {Zurek}(2014)}]{campo2014universality}%
  \BibitemOpen
  \bibfield  {author} {\bibinfo {author} {\bibfnamefont {A.~D.}\ \bibnamefont
  {CAMPO}}\ and\ \bibinfo {author} {\bibfnamefont {W.~H.}\ \bibnamefont
  {Zurek}},\ }\bibfield  {title} {\emph {\bibinfo {title} {Universality of
  phase transition dynamics: Topological defects from symmetry breaking},}\
  }in\ \href@noop {} {\emph {\bibinfo {booktitle} {Symmetry and Fundamental
  Physics: Tom Kibble at 80}}}\ (\bibinfo  {publisher} {World Scientific},\
  \bibinfo {year} {2014})\ pp.\ \bibinfo {pages} {31--87}\BibitemShut {NoStop}%
\bibitem [{\citenamefont {Dziarmaga}(2010)}]{dziarmaga2010dynamics}%
  \BibitemOpen
  \bibfield  {author} {\bibinfo {author} {\bibfnamefont {J.}~\bibnamefont
  {Dziarmaga}},\ }\bibfield  {title} {\emph {\bibinfo {title} {Dynamics of a
  quantum phase transition and relaxation to a steady state},}\ }\href@noop {}
  {\bibfield  {journal} {\bibinfo  {journal} {Advances in Physics}\ }\textbf
  {\bibinfo {volume} {59}},\ \bibinfo {pages} {1063} (\bibinfo {year}
  {2010})}\BibitemShut {NoStop}%
\bibitem [{\citenamefont {Sadler}\ \emph {et~al.}(2006)\citenamefont {Sadler},
  \citenamefont {Higbie}, \citenamefont {Leslie}, \citenamefont
  {Vengalattore},\ and\ \citenamefont {Stamper-Kurn}}]{sadler2006spontaneous}%
  \BibitemOpen
  \bibfield  {author} {\bibinfo {author} {\bibfnamefont {L.}~\bibnamefont
  {Sadler}}, \bibinfo {author} {\bibfnamefont {J.}~\bibnamefont {Higbie}},
  \bibinfo {author} {\bibfnamefont {S.}~\bibnamefont {Leslie}}, \bibinfo
  {author} {\bibfnamefont {M.}~\bibnamefont {Vengalattore}}, \ and\ \bibinfo
  {author} {\bibfnamefont {D.}~\bibnamefont {Stamper-Kurn}},\ }\bibfield
  {title} {\emph {\bibinfo {title} {Spontaneous symmetry breaking in a quenched
  ferromagnetic spinor bose--einstein condensate},}\ }\href@noop {} {\bibfield
  {journal} {\bibinfo  {journal} {Nature}\ }\textbf {\bibinfo {volume} {443}},\
  \bibinfo {pages} {312} (\bibinfo {year} {2006})}\BibitemShut {NoStop}%
\bibitem [{\citenamefont {Chen}\ \emph {et~al.}(2011)\citenamefont {Chen},
  \citenamefont {White}, \citenamefont {Borries},\ and\ \citenamefont
  {DeMarco}}]{chen2011quantum}%
  \BibitemOpen
  \bibfield  {author} {\bibinfo {author} {\bibfnamefont {D.}~\bibnamefont
  {Chen}}, \bibinfo {author} {\bibfnamefont {M.}~\bibnamefont {White}},
  \bibinfo {author} {\bibfnamefont {C.}~\bibnamefont {Borries}}, \ and\
  \bibinfo {author} {\bibfnamefont {B.}~\bibnamefont {DeMarco}},\ }\bibfield
  {title} {\emph {\bibinfo {title} {Quantum quench of an atomic mott
  insulator},}\ }\href@noop {} {\bibfield  {journal} {\bibinfo  {journal}
  {Physical Review Letters}\ }\textbf {\bibinfo {volume} {106}},\ \bibinfo
  {pages} {235304} (\bibinfo {year} {2011})}\BibitemShut {NoStop}%
\bibitem [{\citenamefont {Lamporesi}\ \emph {et~al.}(2013)\citenamefont
  {Lamporesi}, \citenamefont {Donadello}, \citenamefont {Serafini},
  \citenamefont {Dalfovo},\ and\ \citenamefont
  {Ferrari}}]{lamporesi2013spontaneous}%
  \BibitemOpen
  \bibfield  {author} {\bibinfo {author} {\bibfnamefont {G.}~\bibnamefont
  {Lamporesi}}, \bibinfo {author} {\bibfnamefont {S.}~\bibnamefont
  {Donadello}}, \bibinfo {author} {\bibfnamefont {S.}~\bibnamefont {Serafini}},
  \bibinfo {author} {\bibfnamefont {F.}~\bibnamefont {Dalfovo}}, \ and\
  \bibinfo {author} {\bibfnamefont {G.}~\bibnamefont {Ferrari}},\ }\bibfield
  {title} {\emph {\bibinfo {title} {Spontaneous creation of kibble--zurek
  solitons in a bose--einstein condensate},}\ }\href@noop {} {\bibfield
  {journal} {\bibinfo  {journal} {Nature Physics}\ }\textbf {\bibinfo {volume}
  {9}},\ \bibinfo {pages} {656} (\bibinfo {year} {2013})}\BibitemShut {NoStop}%
\bibitem [{\citenamefont {Braun}\ \emph {et~al.}(2015)\citenamefont {Braun},
  \citenamefont {Friesdorf}, \citenamefont {Hodgman}, \citenamefont
  {Schreiber}, \citenamefont {Ronzheimer}, \citenamefont {Riera}, \citenamefont
  {Del~Rey}, \citenamefont {Bloch}, \citenamefont {Eisert},\ and\ \citenamefont
  {Schneider}}]{braun2015emergence}%
  \BibitemOpen
  \bibfield  {author} {\bibinfo {author} {\bibfnamefont {S.}~\bibnamefont
  {Braun}}, \bibinfo {author} {\bibfnamefont {M.}~\bibnamefont {Friesdorf}},
  \bibinfo {author} {\bibfnamefont {S.~S.}\ \bibnamefont {Hodgman}}, \bibinfo
  {author} {\bibfnamefont {M.}~\bibnamefont {Schreiber}}, \bibinfo {author}
  {\bibfnamefont {J.~P.}\ \bibnamefont {Ronzheimer}}, \bibinfo {author}
  {\bibfnamefont {A.}~\bibnamefont {Riera}}, \bibinfo {author} {\bibfnamefont
  {M.}~\bibnamefont {Del~Rey}}, \bibinfo {author} {\bibfnamefont
  {I.}~\bibnamefont {Bloch}}, \bibinfo {author} {\bibfnamefont
  {J.}~\bibnamefont {Eisert}}, \ and\ \bibinfo {author} {\bibfnamefont
  {U.}~\bibnamefont {Schneider}},\ }\bibfield  {title} {\emph {\bibinfo {title}
  {Emergence of coherence and the dynamics of quantum phase transitions},}\
  }\href@noop {} {\bibfield  {journal} {\bibinfo  {journal} {Proceedings of the
  National Academy of Sciences}\ }\textbf {\bibinfo {volume} {112}},\ \bibinfo
  {pages} {3641} (\bibinfo {year} {2015})}\BibitemShut {NoStop}%
\bibitem [{\citenamefont {Navon}\ \emph {et~al.}(2015)\citenamefont {Navon},
  \citenamefont {Gaunt}, \citenamefont {Smith},\ and\ \citenamefont
  {Hadzibabic}}]{navon2015critical}%
  \BibitemOpen
  \bibfield  {author} {\bibinfo {author} {\bibfnamefont {N.}~\bibnamefont
  {Navon}}, \bibinfo {author} {\bibfnamefont {A.~L.}\ \bibnamefont {Gaunt}},
  \bibinfo {author} {\bibfnamefont {R.~P.}\ \bibnamefont {Smith}}, \ and\
  \bibinfo {author} {\bibfnamefont {Z.}~\bibnamefont {Hadzibabic}},\ }\bibfield
   {title} {\emph {\bibinfo {title} {Critical dynamics of spontaneous symmetry
  breaking in a homogeneous bose gas},}\ }\href@noop {} {\bibfield  {journal}
  {\bibinfo  {journal} {Science}\ }\textbf {\bibinfo {volume} {347}},\ \bibinfo
  {pages} {167} (\bibinfo {year} {2015})}\BibitemShut {NoStop}%
\bibitem [{\citenamefont {Anquez}\ \emph {et~al.}(2016)\citenamefont {Anquez},
  \citenamefont {Robbins}, \citenamefont {Bharath}, \citenamefont
  {Boguslawski}, \citenamefont {Hoang},\ and\ \citenamefont
  {Chapman}}]{anquez2016quantum}%
  \BibitemOpen
  \bibfield  {author} {\bibinfo {author} {\bibfnamefont {M.}~\bibnamefont
  {Anquez}}, \bibinfo {author} {\bibfnamefont {B.}~\bibnamefont {Robbins}},
  \bibinfo {author} {\bibfnamefont {H.}~\bibnamefont {Bharath}}, \bibinfo
  {author} {\bibfnamefont {M.}~\bibnamefont {Boguslawski}}, \bibinfo {author}
  {\bibfnamefont {T.}~\bibnamefont {Hoang}}, \ and\ \bibinfo {author}
  {\bibfnamefont {M.}~\bibnamefont {Chapman}},\ }\bibfield  {title} {\emph
  {\bibinfo {title} {Quantum kibble-zurek mechanism in a spin-1 bose-einstein
  condensate},}\ }\href@noop {} {\bibfield  {journal} {\bibinfo  {journal}
  {Physical review letters}\ }\textbf {\bibinfo {volume} {116}},\ \bibinfo
  {pages} {155301} (\bibinfo {year} {2016})}\BibitemShut {NoStop}%
\bibitem [{\citenamefont {Chen}\ \emph {et~al.}(2019)\citenamefont {Chen},
  \citenamefont {Horikoshi}, \citenamefont {Yoshioka},\ and\ \citenamefont
  {Kuwata-Gonokami}}]{chen2019dynamical}%
  \BibitemOpen
  \bibfield  {author} {\bibinfo {author} {\bibfnamefont {Y.}~\bibnamefont
  {Chen}}, \bibinfo {author} {\bibfnamefont {M.}~\bibnamefont {Horikoshi}},
  \bibinfo {author} {\bibfnamefont {K.}~\bibnamefont {Yoshioka}}, \ and\
  \bibinfo {author} {\bibfnamefont {M.}~\bibnamefont {Kuwata-Gonokami}},\
  }\bibfield  {title} {\emph {\bibinfo {title} {Dynamical critical behavior of
  an attractive bose-einstein condensate phase transition},}\ }\href@noop {}
  {\bibfield  {journal} {\bibinfo  {journal} {Physical Review Letters}\
  }\textbf {\bibinfo {volume} {122}},\ \bibinfo {pages} {040406} (\bibinfo
  {year} {2019})}\BibitemShut {NoStop}%
\bibitem [{\citenamefont {Jaksch}\ \emph {et~al.}(1998)\citenamefont {Jaksch},
  \citenamefont {Bruder}, \citenamefont {Cirac}, \citenamefont {Gardiner},\
  and\ \citenamefont {Zoller}}]{jaksch1998cold}%
  \BibitemOpen
  \bibfield  {author} {\bibinfo {author} {\bibfnamefont {D.}~\bibnamefont
  {Jaksch}}, \bibinfo {author} {\bibfnamefont {C.}~\bibnamefont {Bruder}},
  \bibinfo {author} {\bibfnamefont {J.~I.}\ \bibnamefont {Cirac}}, \bibinfo
  {author} {\bibfnamefont {C.~W.}\ \bibnamefont {Gardiner}}, \ and\ \bibinfo
  {author} {\bibfnamefont {P.}~\bibnamefont {Zoller}},\ }\bibfield  {title}
  {\emph {\bibinfo {title} {Cold bosonic atoms in optical lattices},}\
  }\href@noop {} {\bibfield  {journal} {\bibinfo  {journal} {Physical Review
  Letters}\ }\textbf {\bibinfo {volume} {81}},\ \bibinfo {pages} {3108}
  (\bibinfo {year} {1998})}\BibitemShut {NoStop}%
\bibitem [{\citenamefont {Greiner}\ \emph {et~al.}(2002)\citenamefont
  {Greiner}, \citenamefont {Mandel}, \citenamefont {Esslinger}, \citenamefont
  {H{\"a}nsch},\ and\ \citenamefont {Bloch}}]{greiner2002quantum}%
  \BibitemOpen
  \bibfield  {author} {\bibinfo {author} {\bibfnamefont {M.}~\bibnamefont
  {Greiner}}, \bibinfo {author} {\bibfnamefont {O.}~\bibnamefont {Mandel}},
  \bibinfo {author} {\bibfnamefont {T.}~\bibnamefont {Esslinger}}, \bibinfo
  {author} {\bibfnamefont {T.~W.}\ \bibnamefont {H{\"a}nsch}}, \ and\ \bibinfo
  {author} {\bibfnamefont {I.}~\bibnamefont {Bloch}},\ }\bibfield  {title}
  {\emph {\bibinfo {title} {Quantum phase transition from a superfluid to a
  mott insulator in a gas of ultracold atoms},}\ }\href@noop {} {\bibfield
  {journal} {\bibinfo  {journal} {nature}\ }\textbf {\bibinfo {volume} {415}},\
  \bibinfo {pages} {39} (\bibinfo {year} {2002})}\BibitemShut {NoStop}%
\bibitem [{\citenamefont {Cucchietti}\ \emph {et~al.}(2007)\citenamefont
  {Cucchietti}, \citenamefont {Damski}, \citenamefont {Dziarmaga},\ and\
  \citenamefont {Zurek}}]{cucchietti2007dynamics}%
  \BibitemOpen
  \bibfield  {author} {\bibinfo {author} {\bibfnamefont {F.~M.}\ \bibnamefont
  {Cucchietti}}, \bibinfo {author} {\bibfnamefont {B.}~\bibnamefont {Damski}},
  \bibinfo {author} {\bibfnamefont {J.}~\bibnamefont {Dziarmaga}}, \ and\
  \bibinfo {author} {\bibfnamefont {W.~H.}\ \bibnamefont {Zurek}},\ }\bibfield
  {title} {\emph {\bibinfo {title} {Dynamics of the bose-hubbard model:
  Transition from a mott insulator to a superfluid},}\ }\href@noop {}
  {\bibfield  {journal} {\bibinfo  {journal} {Physical Review A}\ }\textbf
  {\bibinfo {volume} {75}},\ \bibinfo {pages} {023603} (\bibinfo {year}
  {2007})}\BibitemShut {NoStop}%
\bibitem [{\citenamefont {Horiguchi}\ \emph {et~al.}(2009)\citenamefont
  {Horiguchi}, \citenamefont {Oka},\ and\ \citenamefont
  {Aoki}}]{horiguchi2009non}%
  \BibitemOpen
  \bibfield  {author} {\bibinfo {author} {\bibfnamefont {N.}~\bibnamefont
  {Horiguchi}}, \bibinfo {author} {\bibfnamefont {T.}~\bibnamefont {Oka}}, \
  and\ \bibinfo {author} {\bibfnamefont {H.}~\bibnamefont {Aoki}},\ }\bibfield
  {title} {\emph {\bibinfo {title} {Non-equilibrium dynamics in
  mott-to-superfluid transition in bose-einstein condensation in optical
  lattices},}\ }in\ \href@noop {} {\emph {\bibinfo {booktitle} {Journal of
  Physics: Conference Series}}},\ Vol.\ \bibinfo {volume} {150}\ (\bibinfo
  {organization} {IOP Publishing},\ \bibinfo {year} {2009})\ p.\ \bibinfo
  {pages} {032007}\BibitemShut {NoStop}%
\bibitem [{\citenamefont {Dziarmaga}\ \emph {et~al.}(2012)\citenamefont
  {Dziarmaga}, \citenamefont {Tylutki},\ and\ \citenamefont
  {Zurek}}]{dziarmaga2012quench}%
  \BibitemOpen
  \bibfield  {author} {\bibinfo {author} {\bibfnamefont {J.}~\bibnamefont
  {Dziarmaga}}, \bibinfo {author} {\bibfnamefont {M.}~\bibnamefont {Tylutki}},
  \ and\ \bibinfo {author} {\bibfnamefont {W.~H.}\ \bibnamefont {Zurek}},\
  }\bibfield  {title} {\emph {\bibinfo {title} {Quench from mott insulator to
  superfluid},}\ }\href@noop {} {\bibfield  {journal} {\bibinfo  {journal}
  {Physical Review B}\ }\textbf {\bibinfo {volume} {86}},\ \bibinfo {pages}
  {144521} (\bibinfo {year} {2012})}\BibitemShut {NoStop}%
\bibitem [{\citenamefont {Shimizu}\ \emph
  {et~al.}(2018{\natexlab{a}})\citenamefont {Shimizu}, \citenamefont {Kuno},
  \citenamefont {Hirano},\ and\ \citenamefont
  {Ichinose}}]{shimizu2018dynamics}%
  \BibitemOpen
  \bibfield  {author} {\bibinfo {author} {\bibfnamefont {K.}~\bibnamefont
  {Shimizu}}, \bibinfo {author} {\bibfnamefont {Y.}~\bibnamefont {Kuno}},
  \bibinfo {author} {\bibfnamefont {T.}~\bibnamefont {Hirano}}, \ and\ \bibinfo
  {author} {\bibfnamefont {I.}~\bibnamefont {Ichinose}},\ }\bibfield  {title}
  {\emph {\bibinfo {title} {Dynamics of a quantum phase transition in the
  bose-hubbard model: Kibble-zurek mechanism and beyond},}\ }\href@noop {}
  {\bibfield  {journal} {\bibinfo  {journal} {Physical Review A}\ }\textbf
  {\bibinfo {volume} {97}},\ \bibinfo {pages} {033626} (\bibinfo {year}
  {2018}{\natexlab{a}})}\BibitemShut {NoStop}%
\bibitem [{\citenamefont {Weiss}\ \emph {et~al.}(2018)\citenamefont {Weiss},
  \citenamefont {Gerster}, \citenamefont {Jaschke}, \citenamefont {Silvi},\
  and\ \citenamefont {Montangero}}]{weiss2018kibble}%
  \BibitemOpen
  \bibfield  {author} {\bibinfo {author} {\bibfnamefont {W.}~\bibnamefont
  {Weiss}}, \bibinfo {author} {\bibfnamefont {M.}~\bibnamefont {Gerster}},
  \bibinfo {author} {\bibfnamefont {D.}~\bibnamefont {Jaschke}}, \bibinfo
  {author} {\bibfnamefont {P.}~\bibnamefont {Silvi}}, \ and\ \bibinfo {author}
  {\bibfnamefont {S.}~\bibnamefont {Montangero}},\ }\bibfield  {title} {\emph
  {\bibinfo {title} {Kibble-zurek scaling of the one-dimensional bose-hubbard
  model at finite temperatures},}\ }\href@noop {} {\bibfield  {journal}
  {\bibinfo  {journal} {Physical Review A}\ }\textbf {\bibinfo {volume} {98}},\
  \bibinfo {pages} {063601} (\bibinfo {year} {2018})}\BibitemShut {NoStop}%
\bibitem [{\citenamefont {Zhou}\ \emph {et~al.}(2020)\citenamefont {Zhou},
  \citenamefont {Li}, \citenamefont {Nath},\ and\ \citenamefont
  {Li}}]{zhou2020quench}%
  \BibitemOpen
  \bibfield  {author} {\bibinfo {author} {\bibfnamefont {Y.}~\bibnamefont
  {Zhou}}, \bibinfo {author} {\bibfnamefont {Y.}~\bibnamefont {Li}}, \bibinfo
  {author} {\bibfnamefont {R.}~\bibnamefont {Nath}}, \ and\ \bibinfo {author}
  {\bibfnamefont {W.}~\bibnamefont {Li}},\ }\bibfield  {title} {\emph {\bibinfo
  {title} {Quench dynamics of rydberg-dressed bosons on two-dimensional square
  lattices},}\ }\href@noop {} {\bibfield  {journal} {\bibinfo  {journal}
  {Physical Review A}\ }\textbf {\bibinfo {volume} {101}},\ \bibinfo {pages}
  {013427} (\bibinfo {year} {2020})}\BibitemShut {NoStop}%
\bibitem [{\citenamefont {Batrouni}\ \emph {et~al.}(2002)\citenamefont
  {Batrouni}, \citenamefont {Rousseau}, \citenamefont {Scalettar},
  \citenamefont {Rigol}, \citenamefont {Muramatsu}, \citenamefont {Denteneer},\
  and\ \citenamefont {Troyer}}]{batrouni2002mott}%
  \BibitemOpen
  \bibfield  {author} {\bibinfo {author} {\bibfnamefont {G.}~\bibnamefont
  {Batrouni}}, \bibinfo {author} {\bibfnamefont {V.}~\bibnamefont {Rousseau}},
  \bibinfo {author} {\bibfnamefont {R.}~\bibnamefont {Scalettar}}, \bibinfo
  {author} {\bibfnamefont {M.}~\bibnamefont {Rigol}}, \bibinfo {author}
  {\bibfnamefont {A.}~\bibnamefont {Muramatsu}}, \bibinfo {author}
  {\bibfnamefont {P.}~\bibnamefont {Denteneer}}, \ and\ \bibinfo {author}
  {\bibfnamefont {M.}~\bibnamefont {Troyer}},\ }\bibfield  {title} {\emph
  {\bibinfo {title} {Mott domains of bosons confined on optical lattices},}\
  }\href@noop {} {\bibfield  {journal} {\bibinfo  {journal} {Physical review
  letters}\ }\textbf {\bibinfo {volume} {89}},\ \bibinfo {pages} {117203}
  (\bibinfo {year} {2002})}\BibitemShut {NoStop}%
\bibitem [{\citenamefont {Hung}\ \emph {et~al.}(2010)\citenamefont {Hung},
  \citenamefont {Zhang}, \citenamefont {Gemelke},\ and\ \citenamefont
  {Chin}}]{hung2010slow}%
  \BibitemOpen
  \bibfield  {author} {\bibinfo {author} {\bibfnamefont {C.-L.}\ \bibnamefont
  {Hung}}, \bibinfo {author} {\bibfnamefont {X.}~\bibnamefont {Zhang}},
  \bibinfo {author} {\bibfnamefont {N.}~\bibnamefont {Gemelke}}, \ and\
  \bibinfo {author} {\bibfnamefont {C.}~\bibnamefont {Chin}},\ }\bibfield
  {title} {\emph {\bibinfo {title} {Slow mass transport and statistical
  evolution of an atomic gas across the superfluid--mott-insulator
  transition},}\ }\href@noop {} {\bibfield  {journal} {\bibinfo  {journal}
  {Physical review letters}\ }\textbf {\bibinfo {volume} {104}},\ \bibinfo
  {pages} {160403} (\bibinfo {year} {2010})}\BibitemShut {NoStop}%
\bibitem [{\citenamefont {Bakr}\ \emph {et~al.}(2010)\citenamefont {Bakr},
  \citenamefont {Peng}, \citenamefont {Tai}, \citenamefont {Ma}, \citenamefont
  {Simon}, \citenamefont {Gillen}, \citenamefont {Foelling}, \citenamefont
  {Pollet},\ and\ \citenamefont {Greiner}}]{bakr2010probing}%
  \BibitemOpen
  \bibfield  {author} {\bibinfo {author} {\bibfnamefont {W.~S.}\ \bibnamefont
  {Bakr}}, \bibinfo {author} {\bibfnamefont {A.}~\bibnamefont {Peng}}, \bibinfo
  {author} {\bibfnamefont {M.~E.}\ \bibnamefont {Tai}}, \bibinfo {author}
  {\bibfnamefont {R.}~\bibnamefont {Ma}}, \bibinfo {author} {\bibfnamefont
  {J.}~\bibnamefont {Simon}}, \bibinfo {author} {\bibfnamefont {J.~I.}\
  \bibnamefont {Gillen}}, \bibinfo {author} {\bibfnamefont {S.}~\bibnamefont
  {Foelling}}, \bibinfo {author} {\bibfnamefont {L.}~\bibnamefont {Pollet}}, \
  and\ \bibinfo {author} {\bibfnamefont {M.}~\bibnamefont {Greiner}},\
  }\bibfield  {title} {\emph {\bibinfo {title} {Probing the
  superfluid--to--mott insulator transition at the single-atom level},}\
  }\href@noop {} {\bibfield  {journal} {\bibinfo  {journal} {Science}\ }\textbf
  {\bibinfo {volume} {329}},\ \bibinfo {pages} {547} (\bibinfo {year}
  {2010})}\BibitemShut {NoStop}%
\bibitem [{\citenamefont {Sherson}\ \emph {et~al.}(2010)\citenamefont
  {Sherson}, \citenamefont {Weitenberg}, \citenamefont {Endres}, \citenamefont
  {Cheneau}, \citenamefont {Bloch},\ and\ \citenamefont
  {Kuhr}}]{sherson2010single}%
  \BibitemOpen
  \bibfield  {author} {\bibinfo {author} {\bibfnamefont {J.~F.}\ \bibnamefont
  {Sherson}}, \bibinfo {author} {\bibfnamefont {C.}~\bibnamefont {Weitenberg}},
  \bibinfo {author} {\bibfnamefont {M.}~\bibnamefont {Endres}}, \bibinfo
  {author} {\bibfnamefont {M.}~\bibnamefont {Cheneau}}, \bibinfo {author}
  {\bibfnamefont {I.}~\bibnamefont {Bloch}}, \ and\ \bibinfo {author}
  {\bibfnamefont {S.}~\bibnamefont {Kuhr}},\ }\bibfield  {title} {\emph
  {\bibinfo {title} {Single-atom-resolved fluorescence imaging of an atomic
  mott insulator},}\ }\href@noop {} {\bibfield  {journal} {\bibinfo  {journal}
  {Nature}\ }\textbf {\bibinfo {volume} {467}},\ \bibinfo {pages} {68}
  (\bibinfo {year} {2010})}\BibitemShut {NoStop}%
\bibitem [{\citenamefont {Bernier}\ \emph {et~al.}(2011)\citenamefont
  {Bernier}, \citenamefont {Roux},\ and\ \citenamefont
  {Kollath}}]{bernier2011slow}%
  \BibitemOpen
  \bibfield  {author} {\bibinfo {author} {\bibfnamefont {J.-S.}\ \bibnamefont
  {Bernier}}, \bibinfo {author} {\bibfnamefont {G.}~\bibnamefont {Roux}}, \
  and\ \bibinfo {author} {\bibfnamefont {C.}~\bibnamefont {Kollath}},\
  }\bibfield  {title} {\emph {\bibinfo {title} {Slow quench dynamics of a
  one-dimensional bose gas confined to an optical lattice},}\ }\href@noop {}
  {\bibfield  {journal} {\bibinfo  {journal} {Physical review letters}\
  }\textbf {\bibinfo {volume} {106}},\ \bibinfo {pages} {200601} (\bibinfo
  {year} {2011})}\BibitemShut {NoStop}%
\bibitem [{\citenamefont {Natu}\ \emph {et~al.}(2011)\citenamefont {Natu},
  \citenamefont {Hazzard},\ and\ \citenamefont {Mueller}}]{natu2011local}%
  \BibitemOpen
  \bibfield  {author} {\bibinfo {author} {\bibfnamefont {S.~S.}\ \bibnamefont
  {Natu}}, \bibinfo {author} {\bibfnamefont {K.~R.}\ \bibnamefont {Hazzard}}, \
  and\ \bibinfo {author} {\bibfnamefont {E.~J.}\ \bibnamefont {Mueller}},\
  }\bibfield  {title} {\emph {\bibinfo {title} {Local versus global
  equilibration near the bosonic mott-insulator--superfluid transition},}\
  }\href@noop {} {\bibfield  {journal} {\bibinfo  {journal} {Physical Review
  Letters}\ }\textbf {\bibinfo {volume} {106}},\ \bibinfo {pages} {125301}
  (\bibinfo {year} {2011})}\BibitemShut {NoStop}%
\bibitem [{\citenamefont {Bernier}\ \emph {et~al.}(2012)\citenamefont
  {Bernier}, \citenamefont {Poletti}, \citenamefont {Barmettler}, \citenamefont
  {Roux},\ and\ \citenamefont {Kollath}}]{bernier2012slow}%
  \BibitemOpen
  \bibfield  {author} {\bibinfo {author} {\bibfnamefont {J.-S.}\ \bibnamefont
  {Bernier}}, \bibinfo {author} {\bibfnamefont {D.}~\bibnamefont {Poletti}},
  \bibinfo {author} {\bibfnamefont {P.}~\bibnamefont {Barmettler}}, \bibinfo
  {author} {\bibfnamefont {G.}~\bibnamefont {Roux}}, \ and\ \bibinfo {author}
  {\bibfnamefont {C.}~\bibnamefont {Kollath}},\ }\bibfield  {title} {\emph
  {\bibinfo {title} {Slow quench dynamics of mott-insulating regions in a
  trapped bose gas},}\ }\href@noop {} {\bibfield  {journal} {\bibinfo
  {journal} {Physical Review A}\ }\textbf {\bibinfo {volume} {85}},\ \bibinfo
  {pages} {033641} (\bibinfo {year} {2012})}\BibitemShut {NoStop}%
\bibitem [{\citenamefont {Del~Campo}\ \emph {et~al.}(2011)\citenamefont
  {Del~Campo}, \citenamefont {Retzker},\ and\ \citenamefont
  {Plenio}}]{del2011inhomogeneous}%
  \BibitemOpen
  \bibfield  {author} {\bibinfo {author} {\bibfnamefont {A.}~\bibnamefont
  {Del~Campo}}, \bibinfo {author} {\bibfnamefont {A.}~\bibnamefont {Retzker}},
  \ and\ \bibinfo {author} {\bibfnamefont {M.~B.}\ \bibnamefont {Plenio}},\
  }\bibfield  {title} {\emph {\bibinfo {title} {The inhomogeneous kibble--zurek
  mechanism: vortex nucleation during bose--einstein condensation},}\
  }\href@noop {} {\bibfield  {journal} {\bibinfo  {journal} {New Journal of
  Physics}\ }\textbf {\bibinfo {volume} {13}},\ \bibinfo {pages} {083022}
  (\bibinfo {year} {2011})}\BibitemShut {NoStop}%
\bibitem [{\citenamefont {G{\'o}mez-Ruiz}\ and\ \citenamefont
  {Del~Campo}(2019)}]{gomez2019universal}%
  \BibitemOpen
  \bibfield  {author} {\bibinfo {author} {\bibfnamefont {F.}~\bibnamefont
  {G{\'o}mez-Ruiz}}\ and\ \bibinfo {author} {\bibfnamefont {A.}~\bibnamefont
  {Del~Campo}},\ }\bibfield  {title} {\emph {\bibinfo {title} {Universal
  dynamics of inhomogeneous quantum phase transitions: suppressing defect
  formation},}\ }\href@noop {} {\bibfield  {journal} {\bibinfo  {journal}
  {Physical Review Letters}\ }\textbf {\bibinfo {volume} {122}},\ \bibinfo
  {pages} {080604} (\bibinfo {year} {2019})}\BibitemShut {NoStop}%
\bibitem [{\citenamefont {Capogrosso-Sansone}\ \emph
  {et~al.}(2008)\citenamefont {Capogrosso-Sansone}, \citenamefont {S{\"o}yler},
  \citenamefont {Prokof’ev},\ and\ \citenamefont
  {Svistunov}}]{capogrosso2008monte}%
  \BibitemOpen
  \bibfield  {author} {\bibinfo {author} {\bibfnamefont {B.}~\bibnamefont
  {Capogrosso-Sansone}}, \bibinfo {author} {\bibfnamefont {{\c{S}}.~G.}\
  \bibnamefont {S{\"o}yler}}, \bibinfo {author} {\bibfnamefont
  {N.}~\bibnamefont {Prokof’ev}}, \ and\ \bibinfo {author} {\bibfnamefont
  {B.}~\bibnamefont {Svistunov}},\ }\bibfield  {title} {\emph {\bibinfo {title}
  {Monte carlo study of the two-dimensional bose-hubbard model},}\ }\href@noop
  {} {\bibfield  {journal} {\bibinfo  {journal} {Physical Review A}\ }\textbf
  {\bibinfo {volume} {77}},\ \bibinfo {pages} {015602} (\bibinfo {year}
  {2008})}\BibitemShut {NoStop}%
\bibitem [{\citenamefont {Sachdev}(2007)}]{sachdev2007quantum}%
  \BibitemOpen
  \bibfield  {author} {\bibinfo {author} {\bibfnamefont {S.}~\bibnamefont
  {Sachdev}},\ }\href@noop {} {\emph {\bibinfo {title} {Quantum phase
  transitions}}}\ (\bibinfo  {publisher} {Wiley Online Library},\ \bibinfo
  {year} {2007})\BibitemShut {NoStop}%
\bibitem [{\citenamefont {Zakrzewski}(2005)}]{zakrzewski2005mean}%
  \BibitemOpen
  \bibfield  {author} {\bibinfo {author} {\bibfnamefont {J.}~\bibnamefont
  {Zakrzewski}},\ }\bibfield  {title} {\emph {\bibinfo {title} {Mean-field
  dynamics of the superfluid-insulator phase transition in a gas of ultracold
  atoms},}\ }\href@noop {} {\bibfield  {journal} {\bibinfo  {journal} {Physical
  Review A}\ }\textbf {\bibinfo {volume} {71}},\ \bibinfo {pages} {043601}
  (\bibinfo {year} {2005})}\BibitemShut {NoStop}%
\bibitem [{\citenamefont {Snoek}\ and\ \citenamefont
  {Hofstetter}(2007)}]{snoek2007two}%
  \BibitemOpen
  \bibfield  {author} {\bibinfo {author} {\bibfnamefont {M.}~\bibnamefont
  {Snoek}}\ and\ \bibinfo {author} {\bibfnamefont {W.}~\bibnamefont
  {Hofstetter}},\ }\bibfield  {title} {\emph {\bibinfo {title} {Two-dimensional
  dynamics of ultracold atoms in optical lattices},}\ }\href@noop {} {\bibfield
   {journal} {\bibinfo  {journal} {Physical Review A}\ }\textbf {\bibinfo
  {volume} {76}},\ \bibinfo {pages} {051603} (\bibinfo {year}
  {2007})}\BibitemShut {NoStop}%
\bibitem [{\citenamefont {Lundh}(2011)}]{lundh2011mott}%
  \BibitemOpen
  \bibfield  {author} {\bibinfo {author} {\bibfnamefont {E.}~\bibnamefont
  {Lundh}},\ }\bibfield  {title} {\emph {\bibinfo {title} {Mott-insulator
  dynamics},}\ }\href@noop {} {\bibfield  {journal} {\bibinfo  {journal}
  {Physical Review A}\ }\textbf {\bibinfo {volume} {84}},\ \bibinfo {pages}
  {033603} (\bibinfo {year} {2011})}\BibitemShut {NoStop}%
\bibitem [{\citenamefont {Krutitsky}\ and\ \citenamefont
  {Navez}(2011)}]{krutitsky2011excitation}%
  \BibitemOpen
  \bibfield  {author} {\bibinfo {author} {\bibfnamefont {K.~V.}\ \bibnamefont
  {Krutitsky}}\ and\ \bibinfo {author} {\bibfnamefont {P.}~\bibnamefont
  {Navez}},\ }\bibfield  {title} {\emph {\bibinfo {title} {Excitation dynamics
  in a lattice bose gas within the time-dependent gutzwiller mean-field
  approach},}\ }\href@noop {} {\bibfield  {journal} {\bibinfo  {journal}
  {Physical Review A}\ }\textbf {\bibinfo {volume} {84}},\ \bibinfo {pages}
  {033602} (\bibinfo {year} {2011})}\BibitemShut {NoStop}%
\bibitem [{\citenamefont {Snoek}(2012)}]{snoek2012collective}%
  \BibitemOpen
  \bibfield  {author} {\bibinfo {author} {\bibfnamefont {M.}~\bibnamefont
  {Snoek}},\ }\bibfield  {title} {\emph {\bibinfo {title} {Collective modes of
  a strongly interacting bose gas: Probing the mott transition},}\ }\href@noop
  {} {\bibfield  {journal} {\bibinfo  {journal} {Physical Review A}\ }\textbf
  {\bibinfo {volume} {85}},\ \bibinfo {pages} {013635} (\bibinfo {year}
  {2012})}\BibitemShut {NoStop}%
\bibitem [{\citenamefont {Rapp}(2013)}]{rapp2013mean}%
  \BibitemOpen
  \bibfield  {author} {\bibinfo {author} {\bibfnamefont {{\'A}.}~\bibnamefont
  {Rapp}},\ }\bibfield  {title} {\emph {\bibinfo {title} {Mean-field dynamics
  to negative absolute temperatures in the bose-hubbard model},}\ }\href@noop
  {} {\bibfield  {journal} {\bibinfo  {journal} {Physical Review A}\ }\textbf
  {\bibinfo {volume} {87}},\ \bibinfo {pages} {043611} (\bibinfo {year}
  {2013})}\BibitemShut {NoStop}%
\bibitem [{\citenamefont {Yan}\ \emph {et~al.}(2017)\citenamefont {Yan},
  \citenamefont {Hui}, \citenamefont {Rigol},\ and\ \citenamefont
  {Scarola}}]{yan2017equilibration}%
  \BibitemOpen
  \bibfield  {author} {\bibinfo {author} {\bibfnamefont {M.}~\bibnamefont
  {Yan}}, \bibinfo {author} {\bibfnamefont {H.-Y.}\ \bibnamefont {Hui}},
  \bibinfo {author} {\bibfnamefont {M.}~\bibnamefont {Rigol}}, \ and\ \bibinfo
  {author} {\bibfnamefont {V.~W.}\ \bibnamefont {Scarola}},\ }\bibfield
  {title} {\emph {\bibinfo {title} {Equilibration dynamics of strongly
  interacting bosons in 2d lattices with disorder},}\ }\href@noop {} {\bibfield
   {journal} {\bibinfo  {journal} {Physical Review Letters}\ }\textbf {\bibinfo
  {volume} {119}},\ \bibinfo {pages} {073002} (\bibinfo {year}
  {2017})}\BibitemShut {NoStop}%
\bibitem [{\citenamefont {Shimizu}\ \emph
  {et~al.}(2018{\natexlab{b}})\citenamefont {Shimizu}, \citenamefont {Hirano},
  \citenamefont {Park}, \citenamefont {Kuno},\ and\ \citenamefont
  {Ichinose}}]{shimizu2018dynamics2}%
  \BibitemOpen
  \bibfield  {author} {\bibinfo {author} {\bibfnamefont {K.}~\bibnamefont
  {Shimizu}}, \bibinfo {author} {\bibfnamefont {T.}~\bibnamefont {Hirano}},
  \bibinfo {author} {\bibfnamefont {J.}~\bibnamefont {Park}}, \bibinfo {author}
  {\bibfnamefont {Y.}~\bibnamefont {Kuno}}, \ and\ \bibinfo {author}
  {\bibfnamefont {I.}~\bibnamefont {Ichinose}},\ }\bibfield  {title} {\emph
  {\bibinfo {title} {Dynamics of first-order quantum phase transitions in
  extended bose--hubbard model: from density wave to superfluid and vice
  versa},}\ }\href@noop {} {\bibfield  {journal} {\bibinfo  {journal} {New
  Journal of Physics}\ }\textbf {\bibinfo {volume} {20}},\ \bibinfo {pages}
  {083006} (\bibinfo {year} {2018}{\natexlab{b}})}\BibitemShut {NoStop}%
\end{thebibliography}%

\end{document}